\documentclass[a4paper,fleqn,usenatbib]{mnras}
 
\usepackage{newtxtext,newtxmath}

\usepackage[T1]{fontenc}
\usepackage{ae,aecompl}

\usepackage{graphicx}	
\usepackage{amsmath}	
\usepackage{amssymb}	

\DeclareMathOperator{\sech}{sech}

\title[Vertical structure of simulated discs in $\Lambda$CDM]{The vertical structure of gaseous galaxy discs in cold dark matter halos}

\author[Ben\'itez-Llambay et al.]{
\parbox[t]{\textwidth}{
       Alejandro Ben\'itez-Llambay$^{1}$\thanks{E-mail: alejandro.b.llambay@durham.ac.uk (ABL)}, Julio
         F. Navarro$^{2}$, Carlos S. Frenk$^{1}$ and \\ Aaron D. Ludlow$^{1,3}$}
\\
\\
$^{1}$Institute for Computational Cosmology, Department of Physics, Durham University, South Road, Durham, DH1 3LE, UK \\
$^{2}$Senior CIfAR Fellow. Department of Physics \& Astronomy, University of Victoria, BC, V8P 5C2, Canada \\
$^{3}$International Centre for Radio Astronomy Research, University of Western Australia, 35 Stirling Highway, \\ Crawley, Western Australia 6009, Australia \
}

\date{Accepted XXX. Received YYY; in original form ZZZ}

\pubyear{2016}

\begin{document}
\label{firstpage}
\pagerange{\pageref{firstpage}--\pageref{lastpage}}
\maketitle

\begin{abstract} 
  We study the vertical structure of polytropic, $P\propto \rho^\Gamma$, centrifugally-supported gaseous discs embedded in cold dark matter (CDM) halos. At fixed radius $R$, the shape of the vertical density profile depends only weakly on whether the disc is self-gravitating (SG) or not (NSG). The disc thickness, set by the midplane sound speed and circular velocity, $(c_s/V_c)R$, in the NSG case, and by the sound speed and surface density, $c_s^2/G\Sigma$, in SG discs, is smaller than either of these scales. SG discs are typically Toomre unstable, NSG discs are stable. Exponential discs in CDM halos with roughly flat circular velocity curves generally ``flare'' outwards. For the polytropic equation of state of the EAGLE simulations, discs whose mass and size match observational constraints are stable (NSG) for $M_d< 3\times 10^9\, M_\odot$ and unstable (SG) at higher masses, if fully gaseous. We test these analytic results using a set of idealized SPH simulations and find excellent agreement. Our results clarify the role of the gravitational softening on the thickness of simulated discs, and on the onset of radial instabilities. EAGLE low-mass discs are non-self-gravitating so the softening plays no role in their vertical structure. High-mass discs, on the other hand, are expected to be self-gravitating and unstable, and may be artificially thickened and stabilized unless gravity is well resolved. Simulations with spatial resolution high enough to not compromise the vertical structure of a disc also resolve the onset of their instabilities, but the converse is not true: resolving instabilities does not guarantee that the vertical structure is resolved. 
\end{abstract}



\section{Introduction}

The vertical structure of centrifugally-supported gaseous discs is a classic astrophysical problem with applications that range from protostellar and protoplanetary discs to spiral galaxies. The physics is well understood: gas discs are systems that result  from energetic losses (``cooling'') and the conservation of angular momentum and whose equilibrium vertical structure is determined by the balance between the effective gas pressure and the vertical compressive force of the gravitational potential. 

Complications arise, however, because the characteristic timescales for the various physical mechanisms at work might differ; because equilibrium might be disturbed by frequent accretion and interaction events; because the gas thermal pressure might be supplemented by bulk and turbulent motions; and, perhaps more importantly in the case of galaxy discs, because the gas may condense into stars that interact with their surrounding gas through energetic feedback processes that may profoundly alter the disc.

Because of these complexities, our understanding of the formation of disc galaxies in a cosmological context,  where systems form hierarchically in a Universe whose matter content is dominated by dark matter (such as the current $\Lambda$ Cold Dark Matter paradigm for structure formation, $\Lambda$CDM), is still incomplete. This is perhaps most apparent in direct hydrodynamical simulations, where early attempts led to simulated discs whose mass and size were quite different from those of their observed counterparts \citep[see; e.g.,][]{Navarro1994,Navarro_Steimetz1997}.

These simulations, however, were useful to diagnose the main shortcomings of those early attempts, most notably insufficient resolution, inefficient feedback, and a far too simplistic modeling of the multiphase, star forming, interstellar medium (ISM) \citep[see; e.g.,][and references therein for an overview of earlier work]{Scannapieco2012}.

Recent improvements on all of these issues have ushered in a new generation of simulations that now reproduce almost routinely the expected mass, size, and scaling laws linking the various structural parameters of galaxy discs ~\citep{Okamoto2005,Governato2007,Brook2011,Guedes2011,Stinson2013,
Hopkins2014,Vogelsberger2014,Schaye2015,
Wang2015,Grand2016,Ferrero2017}.
Despite these success, simulations still struggle to reproduce faithfully the observed vertical structure of discs, and, in particular, of those resembling the thin disc of the Milky Way: with few exceptions, simulated discs tend to be too thick by comparison \citep[see; e.g.,][]{Trayford2017}.

The origin of this discrepancy has not been properly elucidated, but a common suggestion is that limited numerical resolution is the main culprit \citep[e.g.,][]{Governato2004, Grand2016}. In particular, the use of a finite number of particles and softened gravity is often cited but with little quantitative supporting evidence. An additional possibility is that the numerical techniques may be at fault~\citep[see; e.g.,][]{Nelson2006}. Indeed, many simulations utilize particle-based hydrodynamics solvers where discreteness effects and the crude treatment of shocks and discontinuities could, in principle, induce spurious effects that might affect disc scaleheights. 

We examine these issues here using an analytic framework, together with a suite of numerical simulations designed specifically to study the vertical structure of exponential gaseous discs embedded in the gravitational potential of a cold dark matter halo. The simulations use {\tt Gadget-2} \citep{Springel2005}, a code based on the Smoothed Particle Hydrodynamics (SPH) technique, and evolve gaseous discs that settle in a rigid spherical potential modeled after the well-understood mass profile of CDM haloes. Although idealized, our analytic treatment and simulations are nonetheless useful to clarify a number of issues regarding the performance of SPH as well as the importance of the spurious effects introduced by softened gravity and limited mass resolution on the vertical structure of simulated galaxy discs.

We begin in Sec.~\ref{SecAnalRes} with an analytic approach to the problem, which allows us to introduce useful notation and to identify a number of key results that may be used to benchmark the numerical simulations. Sec.~\ref{SecSims} summarizes the key parameters of the code and of the simulation series. Sec.~\ref{SecNumRes} presents our main simulation results and compares them with the analytic expectations. We end by summarizing our main conclusions in Sec.~\ref{SecConc}.

\begin{figure} 
\includegraphics[width=\columnwidth]{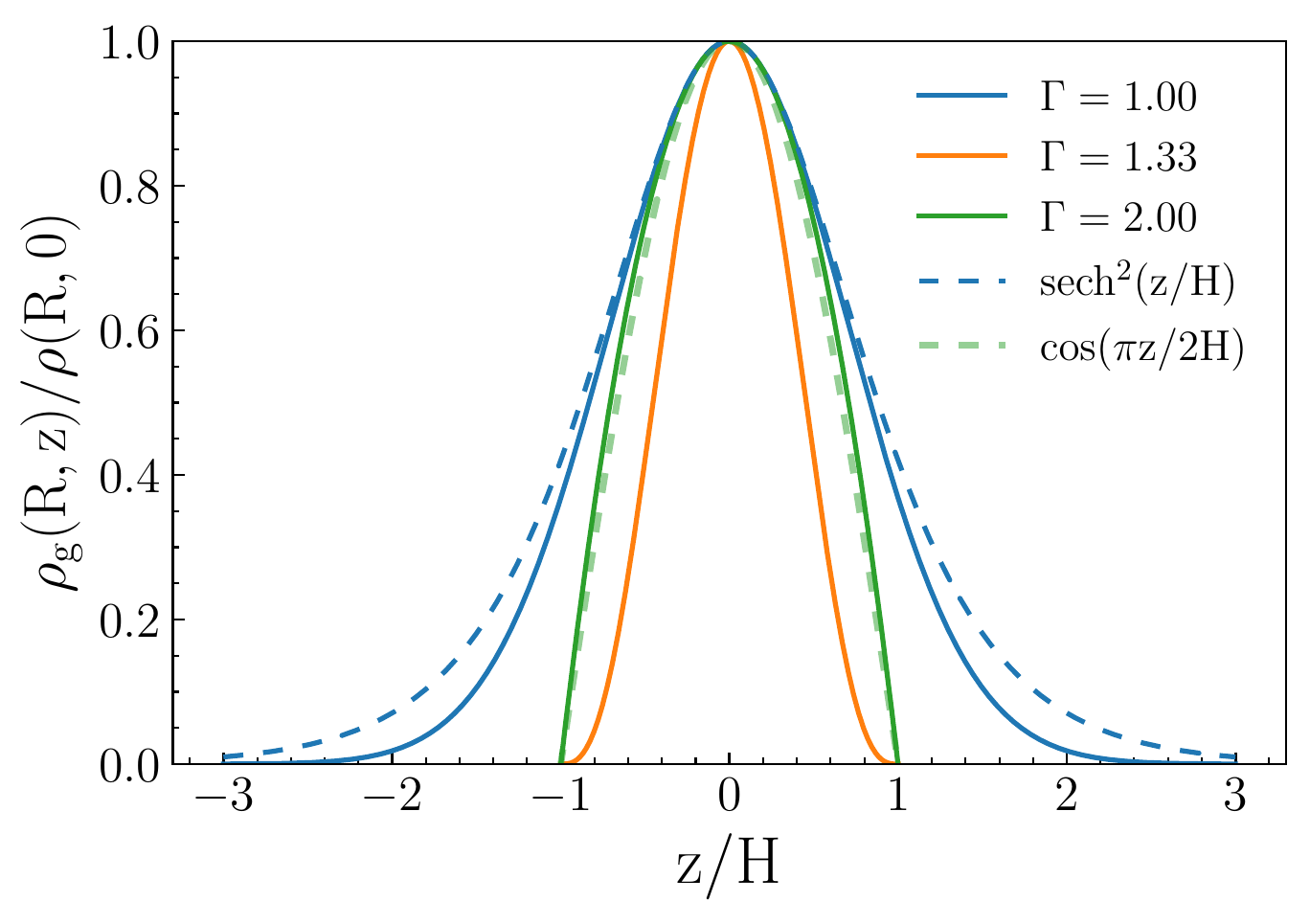} 
\caption{Disc vertical density profiles, as a function of $(z/H)$, for different values of $\Gamma$. Solid and dashed line types correspond to the non-self-gravitating and self-gravitating solutions, respectively. Blue curves correspond to isothermal ($\Gamma=1$) discs; orange to $\Gamma=4/3$; and green to $\Gamma=2$. Note that the shape of the normalized vertical density profile depends weakly on whether the disc is self-gravitating or not, but it is a strong function of the polytropic index, $\Gamma$. Note as well that the height parameter, $H$, has a different physical meaning for different values of $\Gamma$. See text for details.}  \label{FigRhoZ} 
\end{figure}

\begin{figure}
\includegraphics[width=\columnwidth]{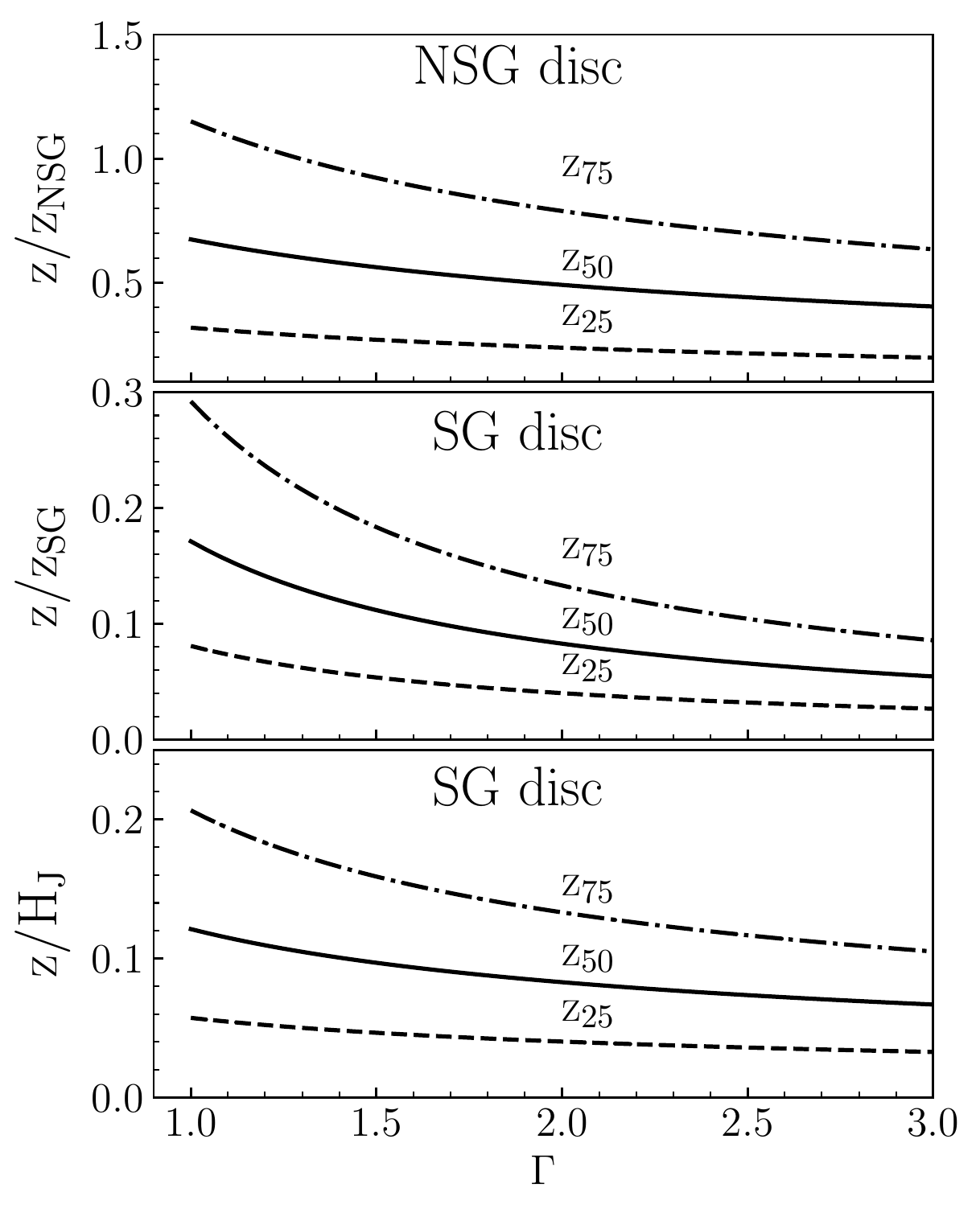}
\caption{The $z$-coordinate enclosing a given fraction (in percentage), $f$, of the disc's column mass, for different values of $\Gamma$. The top panel corresponds to non-self-gravitating discs, with $z_f$ normalized to the ``characteristic'' scaleheight, $z_{\rm NSG}$ (equation~\ref{EqHNSG}). The middle panel is as the top panel but for self-gravitating disks, normalized to  $z_{\rm SG}$ (equation~\ref{EqHSG}). The bottom panel shows the SG scaleheights but normalized to the Jeans length, $H_J$ (equation~\ref{EqHJ}). Note that for a self-gravitating disc the half-mass scaleheight, $z_{50}$, is much smaller than the Jeans length of the system (typically $z_{50}\approx 0.1 \,H_J$). }
\label{FigZGamma}
\end{figure}

\section{Analytic results}
\label{SecAnalRes}

\subsection{Preliminaries}
\label{SecPrelim}

The scaleheight (``thickness'') of a gaseous disc in centrifugal and pressure equilibrium in a dark matter halo is set by the balance between the pressure, $P$, of the gas and the vertical gravity of the disc and the dark matter halo. We present here an order-of-magnitude account of the main parameters that set the aspect ratio of discs and its radial dependence, before presenting, in the following section, a more detailed analysis of polytropic discs that we compare in detail with numerical simulations.  In what follows, we shall assume for simplicity that the disc is fully gaseous, and that the halo is spherical. We use cylindrical coordinates, where $z$ is the rotation axis of the disc, and $R$ is the distance to that axis.

The disc vertical structure is described by the hydrostatic equilibrium equation:
\begin{equation}
\label{Eq:general_hydro_sol}
\displaystyle\frac{1}{\rho_{g}}\displaystyle\frac{{\rm \partial} P}{{\rm \partial} z} = -\displaystyle\frac{{\rm \partial}}{{\rm \partial} z} \left ( \Phi_{g} + \Phi_{\rm dm} \right ),
\end{equation}
where $\rho_{g}$ is the local gas density, and $\Phi_{g}$ and $\Phi_{\rm dm}$ are the gravitational potential of the disc and the dark matter halo, respectively.

In the thin-disc approximation ($z/R << 1$), the contribution of the dark matter halo to the vertical acceleration, at a given point, $(R,z)$, in the space, is given by:
\begin{equation}
\label{Eq:Non-self-gravitating-intro}
 \displaystyle\frac{{\rm \partial} \Phi_{\rm dm}}{{\rm \partial} z} (R,z)= - \displaystyle\frac{V_{\rm dm}^2(R)}{R} \left ( \displaystyle\frac{z}{R} \right ),
\end{equation}
where $V_{\rm dm}^2(R) = G M_{\rm dm}(<R)/R$ and $M_{\rm dm}(<R)$ is the enclosed dark matter mass within a sphere of radius $R$, respectively. In most cases of interest the dark matter dominates the centripetal acceleration, so the circular velocity, $V_c$, is very well approximated by $V_{\rm dm}$; we shall therefore assume that $V_{c}(r)=V_{\rm dm}(r)$ in the remainder of this section.

The contribution of the disc to the vertical component of the gravitational acceleration is obtained by integrating Poisson's equation\footnote{Note that the vertical symmetry of the disc with respect to its midplane ensures that ${\rm \partial} \Phi_{g}/{\rm \partial} z = 0$ at $z=0$.} to yield:
\begin{equation}
\label{Eq:Self-gravitating-intro}
 \displaystyle\frac{{\rm \partial} \Phi_{g}}{{\rm \partial} z}(R,z) =-2\pi G \Sigma(R,z),
\end{equation}
where $\Sigma (R,z)$ is the ($z$-dependent) surface density of the disc, defined by:
\begin{equation}
\label{Eq:Sigma}
\Sigma(R, z) = 2 \displaystyle\int_{0}^z \rho_{g}(R,z') dz'.
\end{equation}

We shall say that a disc is {\it non-self-gravitating} (NSG) if the vertical acceleration profile is primarily described by equation (\ref{Eq:Non-self-gravitating-intro}). In contrast, we will say that a disc is {\it self-gravitating} (SG) if the vertical acceleration profile of the system is primarily described by equation (\ref{Eq:Self-gravitating-intro}). 
Discs of thickness $z_H$, therefore, are self-gravitating when
\begin{equation}
2\pi G \Sigma(R)\gg {V_c^2(R) \over R} {z_H \over R},
\end{equation}
and non-self-gravitating when the inequality is reversed. 

Let us consider first the NSG case, taking for illustration an isothermal disc where pressure and density are linked by a (constant) sound speed, $c_s^2 = (P/\rho_{g})$. As we show below (Sec.~\ref{SecNSG}), in that case, the ``characteristic'' disc scaleheight is
\begin{equation} 
z_{\rm NSG}={c_s\over V_c}R.
\label{EqHNSG}
\end{equation} 
On the other hand, in the SG case, the ``characteristic'' scaleheight is (Sec.~\ref{SecSG}), 
\begin{equation} 
z_{\rm SG}={c_s^2\over G\Sigma(R)}.
\label{EqHSG}
\end{equation}
where $\Sigma(R)$ is the total surface density at radius $R$. 

Note that, in either case, for  galaxy discs where $\Sigma(R)$ decreases with $R$, and where $V_c$ is roughly constant, the disc thickness increases with radius; in other words, most galaxy discs are expected to ``flare'' in the outer regions.

The actual scaleheight of the disc will be the {\it smallest} of those given by equations (\ref{EqHNSG}) and (\ref{EqHSG}) when they differ, or {\it smaller} than either when locally the vertical gravitational pull of the disc and halo are comparable. 

In practice, we find that the scaleheight may be approximated by
\begin{equation}
{1\over z_{\rm H}^2} = {1\over z^2_{\rm NSG}}+{1\over 2\,z_{\rm NSG}z_{\rm SG}}+{1\over z^2_{\rm SG}}. 
\label{EqZh}
\end{equation}

For given scaleheight, the $z$-dependence of the density at $R$, $\rho_{g}(R,z)$, expressed in units of the midplane density, $\rho_g(R,0)$, is not very different for the SG and NSG cases. For example, for an isothermal NSG disc, the density declines exponentially from the midplane ($\rho_{g}(z)\propto \exp{[-(z/z_{\rm H})^2]}$), whereas for an isothermal SG disc we have $\rho_{g}(z) \propto \sech{[-(z/z_{\rm H})^2]}$, which are similar in shape.

\begin{figure}
\includegraphics[width=\columnwidth]{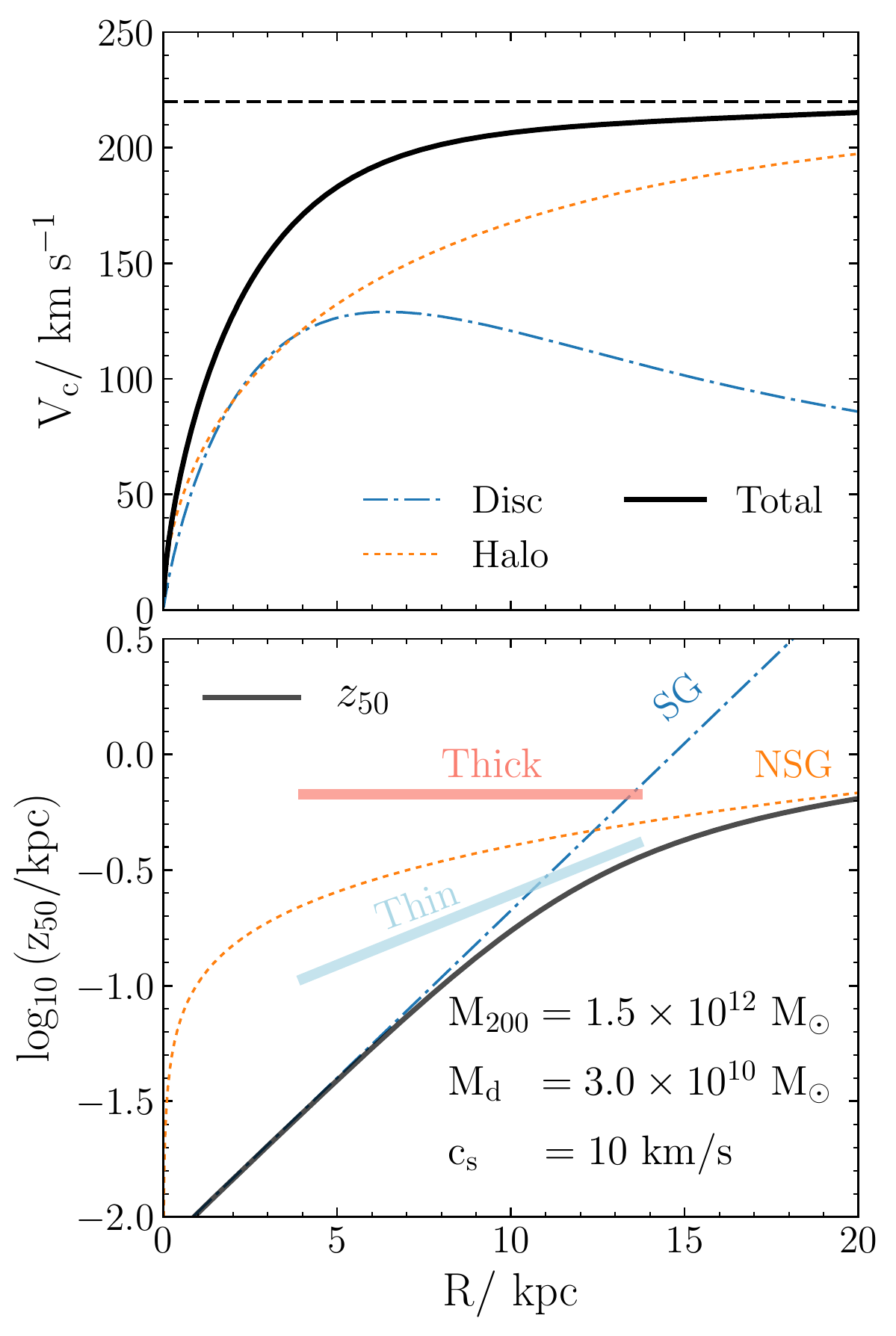}
\caption{{\it Top:} Circular velocity as a function radius for a gaseous isothermal exponential disc of mass $M_{d} \approx 3\times 10^{10} \ M_{\odot}$, scale-radius $R_{d} = 3 \rm \ kpc$, and sound speed $c_{s} = 10 \rm \ km/s$, embedded in a Navarro-Frenk-White spherical halo of virial mass $M_{200} = 1.5 \times 10^{12} \ M_{\odot}$. The parameters of the system have been chosen to roughly approximate those of the low-$\alpha$ (``thin'') disk of the Milky Way \citep{Bovy2016}. {\it Bottom:} Vertical half-mass scaleheight profile of the disk (thick solid curve). Dot-dashed and dotted curves correspond to the results expected for self-gravitating and non-self-gravitating discs, respectively. The thick segments indicate the observational results for the ``thin'' and ``thick'' (high-$\alpha$) disks of the Milky Way, according to \citet{Bovy2016}.
See text for further discussion.}
\label{FigMW}
\end{figure}

Note that SG discs are prone to radial instabilities, which develop when Toomre's parameter, $Q$,  defined by \citep{Toomre1964}:
\begin{equation} 
Q={c_s\kappa \over \pi G \Sigma}
\end{equation} 
is less than a some critical value, $Q_{\rm crit}$, of order unity in the case of infinitesimally thin discs, and about $Q_{\rm crit}\sim 0.6$ for discs of finite thickness \citep[see, e.g.,][]{Wang2010}.
SG discs are generally Toomre unstable, since the epicyclic frequency, $\kappa$, is of order the angular frequency, $\kappa\sim V_c/R$, so that Toomre's  criterion for instability may be rewritten as
\begin{equation} 
G\Sigma >{1\over \pi Q_{\rm crit}} {c_sV_c \over R},
\label{EqToomre}
\end{equation} 
or, equivalently, 
\begin{equation}
{c_s^2 \over G\Sigma}< (\pi Q_{\rm crit})\,  {c_s \over V_c} R,
\end{equation}
which we recognize from equations (\ref{EqHNSG}) and (\ref{EqHSG}) as
\begin{equation}
{z_{\rm SG} \over z_{\rm NSG}} < \pi Q_{\rm crit}.
\end{equation}
The latter condition is generally true for SG discs.

``Self-gravitating'' discs are, therefore, almost always unstable\footnote{We refer the reader to the App.~\ref{App:Instabilities}, where we show that there is a narrow range of parameters for which discs can be self-gravitating and stable.}. Conversely, it is straightforward to show that NSG discs are, in general, Toomre stable.

The above discussion shows that the thickness of a gaseous disc and its stability are governed, at a given radius, by the combination of $c_s$, $V_c$, and $\Sigma$ (for a fixed equation of state). It is illustrative to consider the simple case of an exponential, isothermal disc with a ``flat'' rotation curve (i.e., constant circular velocity). In this case, $z_{\rm NSG}/R$ is a constant. $z_{\rm SG}/R$, on the other hand, diverges at small and large radii and has a minimum at the exponential scale radius, $R_d$. Therefore, when $(c_s/V_c)R_d< c_s^2/G \Sigma(R_d)$ , or, equivalently, when
\begin{equation}
{G \, \Sigma(R_d) \, R_d \over c_s \,V_c(R_d)}<1 
\label{EqInstCond}
\end{equation}
the disc will be NSG and, consequently, stable everywhere. If, on the other hand, $z_{\rm SG}<z_{\rm NSG}$ at $R_d$, then there will be a region around $R_d$ where the disc will likely be Toomre unstable.

Radial instabilities introduce an additional scale in the problem, namely the characteristic size (or mass) of the clumps that first develop. This is well approximated by the ``critical'' wavelength that results from the linear stability analysis of differentially rotating discs \citep[see, e.g., Sec. 6.2.3 of][]{Binney2008},
\begin{equation}
\lambda_{\rm crit}=\displaystyle\frac{4 \pi^2 G \Sigma}{\kappa^2}.
\label{EqLambdaCritDef}
\end{equation}
This wavelength usually exceeds the disc scaleheight by a fairly large factor. Indeed, for SG discs
\begin{equation}
\displaystyle\frac{z_{\rm SG}}{\lambda_{\rm crit}} \propto \displaystyle\frac{c_s^2 V_c^2}{(2\pi G\Sigma R)^2} << 1,
\end{equation}
where the latter inequality follows from equation (\ref{EqToomre}).  This implies that numerical simulations with spatial resolution adequate enough to resolve the scaleheight of a SG disc also resolve the onset of radial instabilities. The converse, however, is not necessarily true.

NSG discs, on the other hand, are generally stable, and have scaleheights determined by the halo rather than by the disc, placing much less stringent constraints on the spatial resolution needed to simulate their evolution.

Finally, it is interesting to compare this instability scale with the ``Jean's length'', $H_J$, often used in the literature and defined by
\begin{equation} 
H_{J}^2 = {\pi c_{s}^2 \over G\rho_{g}}.  
\label{EqHJ}
\end{equation} 
It is straightforward to show that $\lambda_{\rm crit}$ is generally much greater than $H_J$. Indeed, for SG discs we have that 
\begin{equation}
{\lambda_{\rm crit} \over H_{J}} \sim \left( {c_s R \over z_H V_c} \right)^{2} \sim  \left( {z_{\rm NSG} \over z_H}\right)^2 \gg 1,
\label{EqLcritHjSG}
\end{equation}
where we have used $z_H\approx z_{\rm SG}=c_s^2/G\Sigma$ (equation~\ref{EqHSG}) and that  the epicyclic frequency is of order of the angular frequency; $\kappa\sim V_c/R$.

This implies that the spatial resolution required to follow the onset of radial instabilities in the disc are much less stringent than those required to resolve the Jeans length. In other words, we expect SG discs to become  unstable even when the numerical resolution is too poor to properly resolve the true vertical height, or when the number of particles is too small to properly resolve the Jeans length. The main requirement for such instabilities to grow is that self-gravity be faithfully approximated on scales smaller than $\lambda_{\rm crit}$, which places an upper value on the gravitational softening used in simulations that attempt to resolve them.

Instabilities play an important role in numerical simulations like the ones we describe below. Stable discs transform gas into stars roughly uniformly throughout the disc but unstable discs break up into self-bound clumps before turning into stars. The subsequent evolution of these stellar clumps may play an important role in setting the vertical structure of simulated galaxy discs, an issue to which we return in Sec~\ref{SecStellarD}.

\subsection{Polytropic discs}
\label{SecPoly}

Following the approach of Sec.~\ref{SecPrelim}, we consider separately the non-self-gravitating and self-gravitating cases before deriving a simple formula that approximates well the general solution for the scaleheight of a gaseous disc in a spherical dark matter halo. Our solutions apply to a polytropic equation of state (EoS) for the gas, of the form,
\begin{equation}
\label{EqEoS}
P = P_{\rm eos} \left ( \displaystyle\frac{\rho_{g}}{\rho_{\rm eos}} \right )^{\Gamma} = \displaystyle\frac{c_{s}^2\rho_{g}}{\Gamma}
\end{equation}
where $P_{\rm eos}$ and $\rho_{\rm eos}$ determine the normalization of the relation, $\Gamma$ is the polytropic index, and $c_{s}=\left ( \partial P / \partial \rho \right )^{1/2}$ is the sound speed. Isothermal discs ($\Gamma=1$) have constant sound speed, but $c_s$ increases with density for $\Gamma>1$. 

A common assumption in cosmological numerical simulations such as those of the EAGLE project \citep{Schaye2015, Crain2015}, is to adopt $\Gamma=4/3$, $P_{\rm eos} \sim 1.10 \rm \ g \ cm^{-1} \ s^{2}$, and $\rho_{\rm eos}/m_p=0.1 \rm \ cm^{-3}$, where $m_{p}$ is the proton mass, which gives
\begin{equation}
c_s=c_{\rm s,eos} \left ( \displaystyle\frac{\rho_{g}}{\rho_{\rm eos}} \right )^{(\Gamma-1)/2},
\label{EqCsEoS}
\end{equation}
with $c_{\rm s,eos} = (P_{\rm eos} \Gamma/\rho_{\rm eos})^{1/2} \sim 9.4 \rm \ km/s$. We will refer to this equation state as the EAGLE EoS. 

The choice of $\Gamma=4/3$ is motivated by the fact that the Jeans mass scale, $M_J\propto \rho_{g} H_J^3\propto \rho_{g}^{-2+3\Gamma/2}$, becomes independent of density for that polytropic index, and is implemented to prevent spurious fragmentation due to finite numerical resolution as unstable clumps develop in the disc~\citep[][]{Bate1997,Schaye2008}.

\subsubsection{Non-self-gravitating (NSG) discs}
\label{SecNSG}

Recalling that the vertical structure of thin ($z\ll R$) NSG discs is set by the balance between the pressure gradient and the vertical acceleration profile of the halo, we have in the NSG regime that
\begin{equation}
\label{Eq:HydroEq}
\displaystyle\frac{1}{\rho_{g}} \displaystyle\frac{{\rm \partial}P}{{\rm \partial}z} = -\displaystyle\frac{V_{c}^2(R)}{R} \left ( \displaystyle\frac{z}{R} \right ).
\end{equation}
Inserting equation (\ref{EqEoS}) into equation (\ref{Eq:HydroEq}), we can solve for the vertical density profile of the disc:
\begin{equation}
\label{EqNSGRhoZ}
  \displaystyle\frac{\rho_{g}(R,z)}{\rho_{g}(R,0)} = \begin{cases}
     \left [ 1 - (z/H_{\rm NSG})^2   \right ]^{1/(\Gamma-1)}, &\mbox{if } \Gamma \ne 1 \\
        \exp [- (z/H_{\rm NSG})^2], &\mbox{if } \Gamma = 1\\
  \end{cases}
\end{equation}
where the non-self-gravitating height parameter is defined as
\begin{equation}
\label{Eq:Height-nograv}
  H_{\rm NSG}= \alpha(\Gamma) (c_{s,0}/V_c) R,
\end{equation}
$c_{s,0}$ is the midplane sound speed, and $\alpha({\Gamma})$ is given by
\begin{equation}
\label{Eq:alpha_gamma}
  \alpha({\Gamma})= \begin{cases} 
	 \sqrt{\displaystyle\frac{2}{\Gamma-1}}  &\mbox{if } \Gamma \ne 1 \\
         \sqrt{2} ,  &\mbox{if } \Gamma = 1 .\\
  \end{cases}
\end{equation}

The solid lines in Fig.~\ref{FigRhoZ} show the resulting $z$-dependence of the density profile, for various values of $\Gamma$.  The height parameter $H_{\rm NSG}$ defined by equation (\ref{Eq:Height-nograv}) is numerically convenient, but has different meanings for different values of $\Gamma$. A more physically meaningful definition of scaleheight is provided by the value of $z_f$ that contains a given fraction, $f$, of the disc column mass at each radius.

We show this in the top panel of Fig.~\ref{FigZGamma}, where we plot, in units of the ``characteristic'' NSG scaleheight, $z_{\rm NSG}$ (equation~\ref{EqHNSG}), the heights containing various fractions of the disc column mass as a function of $\Gamma$.

\begin{table}
 \caption{Values of $F_c$ (equation~\ref{Eq:Correction_factor}) and $y_f$ (equation~\ref{Eq:zf}), for different values of the polytropic index, $\Gamma$. Use these numbers to convert the scaleheight parameter, $H$, into "characteristic" scaleheight values, such as the half-mass scaleheight, $z_{50} = y_{50} H$. }
 \label{Table:zf}
 \begin{tabular}{cllll}
 \hline
 $\Gamma$ & $F_c$ & $y_{25}$ & $y_{50}$ & $y_{75}$ \\
 \hline
 1   & 0.886 & 0.225 & 0.477 & 0.813 \\
 4/3 & 0.457 & 0.116 & 0.242 & 0.402 \\
 2   & 2/3   & 0.168 & 0.347 & 0.558 \\
 \hline
 \end{tabular}
\end{table}

Finally, we quote the relation between midplane density and surface density, which is useful to derive $z_{f}$. Writing $\rho_{g}(R,z) = \rho_{g}(R,0) \, g(z/H_{NSG},\Gamma)$, we have that 
\begin{equation}
\label{Eq:rho_to_sigma}
\Sigma(R) = 2 F_c\, \rho_{g}(R,0) \, H_{\rm NSG} 
\end{equation}
where $F_c$ is the integral of $g(z/H_{NSG},\Gamma)$:
\begin{equation}
\label{Eq:Correction_factor}
F_c = \begin{cases}
	    \displaystyle\int_{0}^{1} \left ( 1 - u^2 \right )^{1/(\Gamma-1)} du & \Gamma \ne 1 \\
	    \sqrt{\pi}/2 & \Gamma = 1. \\
  \end{cases} 
\end{equation}
For $\Gamma = 4/3$, $F_{c} \sim 0.4$, indicating that the vertical density profile of a NSG disc is poorly  approximated by a top-hat model in which $\Sigma(R) = 2 \rho_{g} (R,0) H_{\rm NSG}$ (see also Fig.~\ref{FigRhoZ}). 

The scaleheight $z_{f}$ is simply given by 
\begin{equation} 
z_{f} = y_f\,H_{NSG}, 
\label{Eq:zf} 
\end{equation}
where $y_{f}$ is given by
\begin{equation}
\label{Eq:yf}
 \displaystyle\int_{0}^{y_f} g(u,\Gamma) \ du = f\,F_{c}.
\end{equation}
We tabulate various values of $y_{f}$, for different values of $\Gamma$ in Table~\ref{Table:zf}.\footnote{Although equation (\ref{Eq:zf}) is strictly valid in the NSG regime, we show in Sec.~\ref{SecSG} that it can also be applied for self-gravitating discs.} 

The analysis above shows that polytropic NSG discs with $\Gamma>1$ have well defined maximum heights (given by equation~\ref{Eq:Height-nograv}), where the density vanishes. Isothermal discs, on the other hand, extend to arbitrarily large heights.

\begin{figure}
\includegraphics[width=\columnwidth]{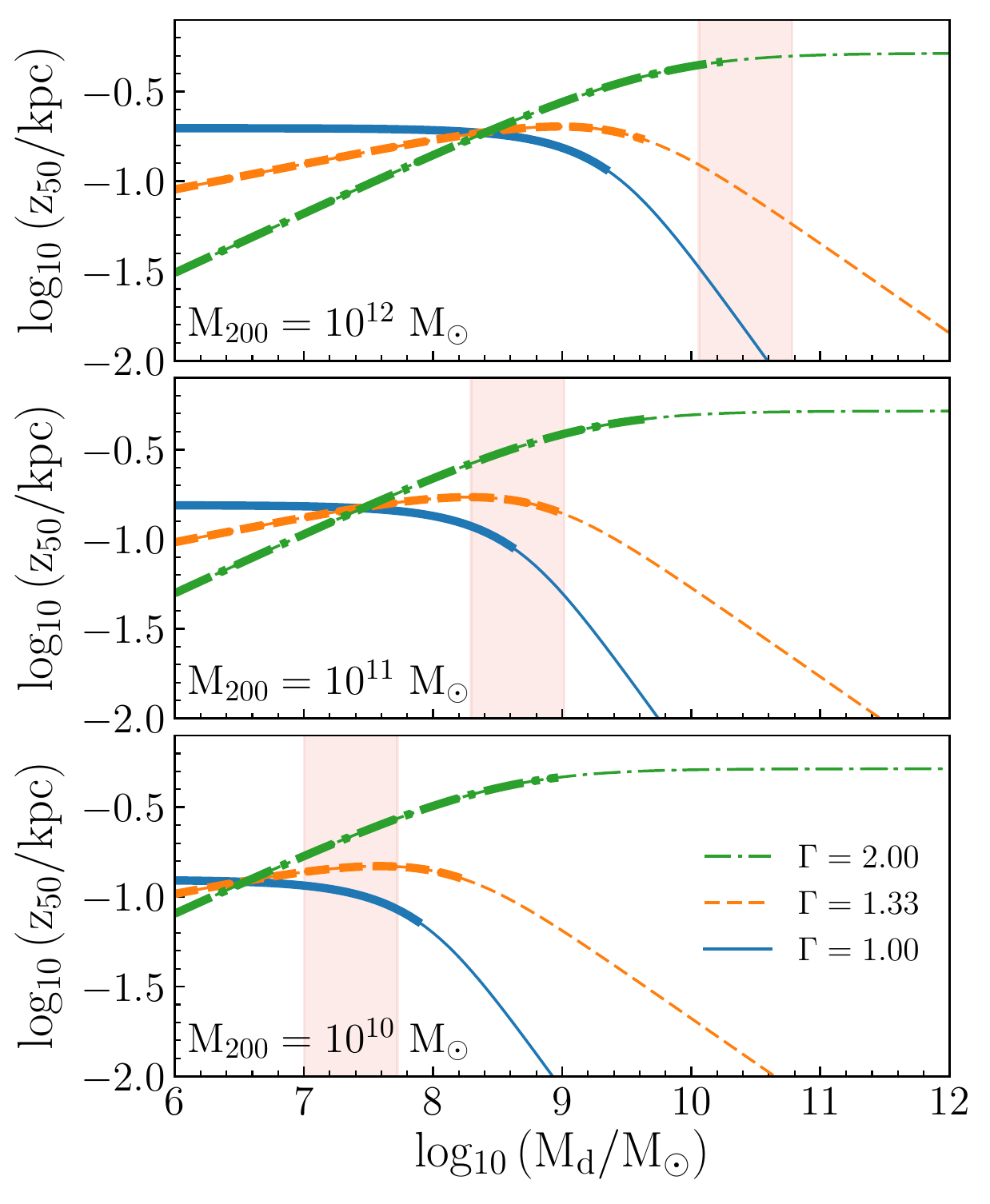}
\caption{Half-mass scaleheight, $z_{50}$, as a function of disk mass, for three different halo masses, as labelled in each panel, and for three different polytropic indexes, namely, $\Gamma = 1, \Gamma = 4/3$, and $\Gamma = 2$. Discs are assumed to have the same scale radius in each panel, namely $R_d=3.0$, $1.10$, and $0.44$ kpc from top to bottom. Red shaded regions show the acceptable range of disc masses, at a given halo mass, according to the ``abundance-matching'' model of \citet{Behroozi2013}, with a scatter of a factor of $2$ in galaxy mass. Line types switch  from thick to thin when discs become Toomre unstable. Note that ``realistic'' disc galaxies that form in halos less massive than $\sim 10^{11} \ M_{\odot}$ are expected to be stable. }
\label{FigMdz50}
\end{figure}

\subsubsection{Self-gravitating (SG) discs}
\label{SecSG}

We can derive the scaleheight of a polytropic self-gravitating disc in a simple (although approximate\footnote{More detailed treatments may be found in, for example, ~\citet[][]{Spitzer1942, Goldreich1965}, and App.~\ref{App:Appendix_density_profiles}.}) way.
The vertical structure of SG discs is set by the balance between its pressure and its own vertical gravity:
\begin{equation}
\label{Eq:HydroSelf}
\displaystyle\frac{1}{\rho_{g}} \displaystyle\frac{{\rm \partial} P}{{\rm \partial} z} = - 2\pi G \Sigma(R,z),
\end{equation}

Equation (\ref{Eq:HydroSelf}) looks simpler expressed in terms of the disc's surface density only:
\begin{equation}
\displaystyle\frac{{\rm \partial} P}{{\rm \partial} \Sigma} = -\pi G \Sigma (R,z),
\end{equation}
in which we used the fact that, by definition, ${\rm \partial} \Sigma / {\rm \partial} z$ = 2$\rho_{g}$. Integration of the previous equation yields:
\begin{equation}
\label{Eq:Pressure-selfgrav}
\displaystyle\frac{P(R,z)}{P(R,0)} = \left [ 1-\displaystyle\frac{\pi G \Sigma^2(R,z)}{2P(R,0)} \right ],
\end{equation}
with $P(R,0)$ being the midplane pressure of the disc.

Similarly to equation (\ref{EqNSGRhoZ}), equation (\ref{Eq:Pressure-selfgrav}) shows that the pressure of the disc (and therefore its density) vanishes at:
\begin{equation}
\label{Eq:Sigma_selfgrav}
\Sigma(R,H_{\rm SG})=\Sigma(R)= \left ( \displaystyle\frac{2 P(R,0)}{\pi G} \right )^{1/2},
\end{equation}
which implicitly defines the height parameter, $H_{\rm SG}$, of a SG polytropic disc. The value of $H_{\rm SG}$ can be found by combining equations (\ref{Eq:Sigma_selfgrav}) and (\ref{Eq:Sigma}):
\begin{equation}
\label{Eq:Height_grav}
 H_{\rm SG} = \left ( \displaystyle\frac{1}{\pi \Gamma F_c} \right ) \left ( \displaystyle\frac{c_{s,0}^2}{G \Sigma(R)} \right ),
\end{equation}
where  we have assumed that the vertical dependence of the density profile can be approximated by equation (\ref{EqNSGRhoZ}), so that the $F_c$ factor (equation~\ref{Eq:Correction_factor}) is the same as derived for the NSG regime. Although the vertical dependence of the density in a SG disc differs form that of a NSG disc, in practice the differences are quite small.

Analytic forms for the SG density profile may be computed for some values of $\Gamma$ \citep[see][and App.~\ref{App:Appendix_density_profiles}]{Goldreich1965}, and can be derived numerically for other values. In particular,
\begin{equation}
\label{App:Density_selfgrav}
\displaystyle\frac{\rho_{g}(R,z)}{\rho_{g}(R,0)} = \begin{cases}
\sech^2 \left ( z/H_{\rm SG} \right ),  &\mbox{if } 	\Gamma =1 \\
\cos \left ( z/H_{\rm SG} \right ), &\mbox{if } \Gamma = 2. \\
\end{cases}
\end{equation}
For these values of $\Gamma$, we have that
\begin{equation}
\label{App:heights}
H_{\rm SG} = \begin{cases}
\left ( \displaystyle{c_{s,0}^2/2\pi G \rho_{g}(R,0)} \right )^{1/2}, &\mbox{if } \Gamma = 1. \\
\displaystyle ({\pi/2}) \left ( \displaystyle {P_{\rm eos}/2\pi G \rho_{\rm eos}^2} \right )^{1/2}, &\mbox{if } \Gamma = 2. \\
\end{cases}
\end{equation}

We compare SG and NSG density profiles in Fig.~\ref{FigRhoZ}. This figure confirms that, for given $\Gamma$, the $z$-dependence of the density (scaled to the midplane density and the scaleheight, $H$) is very similar for SG and NSG discs. Self-gravitating discs thus differ from their non-self-gravitating counterparts mostly in the value of their scaleheights, and not in the shape of their vertical density profile. This implies that equation (\ref{Eq:zf}) (or values quoted in Table~\ref{Table:zf}) can be used to calculate different column mass scaleheights, $z_{f}$, even when discs are self-gravitating.

Various scaleheights, $z_f$, of SG discs, expressed in units of their ``characteristic'' value, $z_{\rm SG}$ (equation~\ref{EqHSG}), are shown as a function of $\Gamma$ in the middle panel of Fig.~\ref{FigZGamma}.

We conclude by comparing the characteristic heights of SG discs with the ``Jeans scaleheight'', $H_J$, in the bottom panel of Fig.~\ref{FigZGamma}. As expected, the disc characteristic thickness is much smaller than $H_J$ and, when expressed in units of $H_J$, is only a weak function of $\Gamma$.

\begin{figure}
\includegraphics[width=\columnwidth]{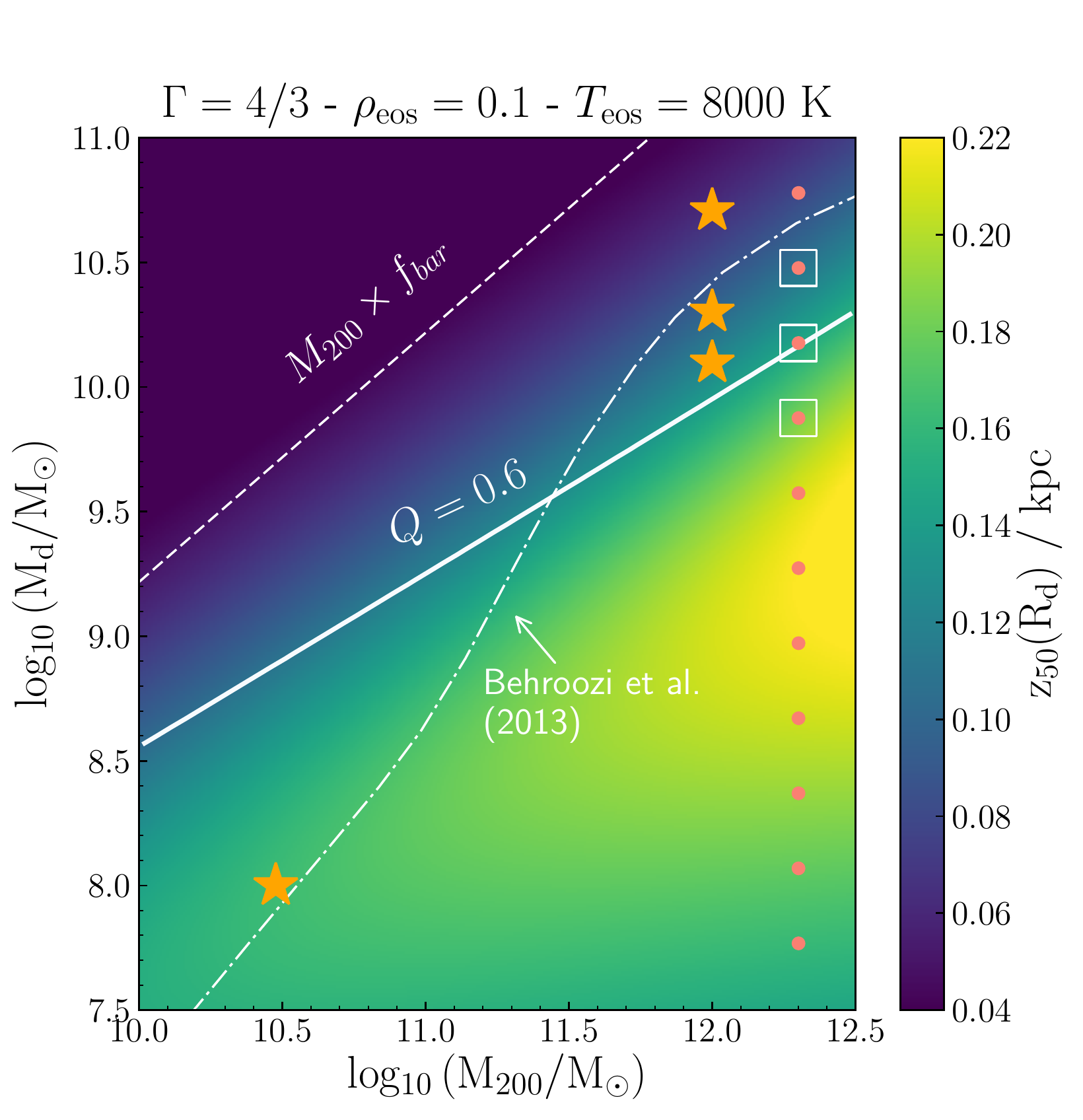}
\caption{Disk mass vs virial mass, colored by the half-mass scaleheight, $z_{50}(R_d)$, for $\Gamma = 4/3$ exponential discs, modeled with a polytropic EAGLE EoS. The scale radius of discs is fixed for each $M_{200}$ and chosen so that the half-mass radius $R_{h} = 1.678\, R_{d}=0.2\, r_s$, where $r_s$ is the NFW scale radius of the abundance-matching halo given by the model \citet{Behroozi2013}.  Top area (shaded in dark blue) indicates masses that exceed the total baryonic budget of the halo, ($\Omega_{\rm bar}/\Omega_{\rm m}) M_{200}$, and are excluded in CDM models. Dot-dashed white line tracks the halo mass-galaxy mass relation given by the abundance matching model of~\citet{Behroozi2013}. Solid thick line shows the neutral stability region where the Toomre parameter is $Q = Q_{\rm crit}=0.6$; for our choice of $R_h$, exponential discs above this line are Toomre unstable. Filled red circles indicate the systems we discuss in Section~\ref{SecNumRes}. White squares correspond to systems shown in Fig.~\ref{FigDiscs}. Orange stars highlight the parent gas discs of the numerical experiments discussed in Section~\ref{SecStellarD}. }
\label{FigMbarMh}
\end{figure}

\subsubsection{A general solution for the disc scaleheight}
\label{SecH}

When the contributions of both disc and halo are important for the vertical acceleration, the true scaleheight of the disc must be computed numerically, by solving 
\begin{equation}
{1 \over \rho_g}{\partial P \over \partial z}=-{\partial \over \partial z} {(\Phi_{\rm h}+\Phi_g)}.
\label{EqdPdzh+d}
\end{equation}
This requires iterative procedures, since the disc potential and sound speed depend on its thickness, which is what we are trying to compute in the first place. 

For an isothermal disc ($\Gamma = 1$), it is sufficient to use the circular velocity, midplane sound speed, and surface density to compute the scaleheight parameters $H_{\rm NSG}$ (Eq.~\ref{Eq:Height-nograv}) and $H_{\rm SG}$ (Eq.~\ref{Eq:Height_grav}). We may use equation (\ref{Eq:zf}) to estimate the half-mass scaleheight, $z_{50}$, corresponding to each case. Then the actual half-mass scaleheight may be approximated by the square harmonic mean (equation \ref{EqZh}), which proves reasonably accurate for most practical purposes\footnote{We refer the reader to App.~\ref{App:general_scaleheight} for a derivation of this formula.}. 

For a polytropic disc ($\Gamma > 1$), we can use the same approximation, but the midplane sound speed of the system, which depends on the midplane density of the disc, must be found self-consistently. A practical solution involves the following procedure. For an initial (arbitrary) guess of $H$, we use equation (\ref{Eq:rho_to_sigma}) to link the disc surface density to the actual midplane density, which is used to calculate the midplane sound speed; the resulting sound speed is then used to calculate $H_{SG}$ (equation~\ref{Eq:Height_grav}) and $H_{NSG}$ (equation~\ref{Eq:Height-nograv}) self-consistently, which are then inserted into equation (\ref{EqZh}) to obtain a new value of $H$; we iterate this procedure until convergence is reached. Finally, the half-mass scaleheight may be calculated using equation (\ref{Eq:zf}), or the data provided in Table~\ref{Table:zf}.

\subsubsection{The solar circle as illustrative example}

For illustration, let us compute a few characteristic values for the Milky Way (MW) disc at the solar circle, $R_{0}=8$ kpc, using  the $\Gamma=4/3$ fiducial EAGLE EoS. Assuming that the Milky Way's midplane density (the ``Oort limit'')  is $\rho_g(R_0,0)=\rho_{\rm OL}\sim 0.04 \, M_\odot/$pc$^3$~\citep{Bovy2017}, we find an effective midplane sound speed of $\sim 15$ km/s, and a fiducial  NSG scaleheight parameter $H_{\rm NSG}=\alpha(\Gamma) (c_{s,0}/V_{c}(R_{0}))\,R_0\sim 1.0$ kpc, where we assumed $V_{c}(R_{0})\sim 220 \rm \ km/s$. Assuming a surface density $\Sigma(R_0)\sim 40 \, M_\odot/$pc$^2$~\citep{Kuijken1989,Bovy2013}, its fiducial SG scaleheight parameter would be $H_{\rm SG}=(\pi \Gamma F_{c})^{-1}\,c_{s,0}^2/G\Sigma(R_0)\sim 726$ pc. 

The respective half-mass scaleheights would be $320$ pc and $180$ pc for the NSG and SG cases, respectively, for an actual scaleheight of $z_{50}(R_0)\sim 150$ pc, computed using equation (\ref{EqZh}). Thus, gaseous Milky Way-like discs in EAGLE are expected to be  dominated by their own self-gravity at radii comparable to the solar circle. 

\subsection{Exponential discs in CDM halos: the Milky Way}
\label{SecMW}

We apply the above results to a worked example, where we consider an exponential gaseous disc embedded in a cold dark matter halo\footnote{In Sec.~\ref{SecStellarD} we show that similar behaviour is expected for stellar discs, at least in a simple model.}. The halo is modeled by a Navarro-Frenk-White ~\citep[hereafter NFW;][]{Navarro1996,Navarro1997} rigid potential. We choose model parameters inspired by the structure of the Milky Way disc, as reported by \citet{Bovy2016}, although of course, the disk of the Milky Way is primarily stellar rather than gaseous. These authors distinguish two separate disc components according to the abundance of $\alpha$ elements; an $\alpha$-poor (``thin'') component that includes Sun-like stars, and an $\alpha$-enhanced (``thick'') disc.

The thin disc has mass $M_d\approx 3\times 10^{10} \, M_\odot$ and exponential scale length $R_d=3$ kpc. If placed in an NFW halo with virial\footnote{Virial quantities correspond to those of the sphere where the enclosed mean density is $200$ times the critical density for closure, $\rho_{\rm crit}=3H_0^2/8\pi G = 2.775 \times 10^{11} h^{2} M_{\odot}/\rm Mpc^3$, and are identified with a $200$ subscript. Throughout this paper, we assume $h = 0.7$.}  mass $M_{200}=1.5 \times 10^{12}\, M_\odot$ and concentration $c=8$ (equation \ref{EqVNFW}), it would have a reasonably flat circular velocity curve that peaks at roughly $220$ km/s (see top panel of Fig.~\ref{FigMW}). The same panel shows the contribution of the disc to the circular velocity, as well as that of the halo. 

The bottom panel of Fig.~\ref{FigMW} shows, as a function of $R$, the two half-mass scaleheight profiles SG (dot-dashed)  and NSG (dashed) for gaseous discs, assuming, for simplicity, an isothermal EoS ($\Gamma = 1$ and $c_s=10$ km/s). The solid thick line is the result for the ``true'' profile, obtained from equation (\ref{EqZh}). 

As stated in Sec~\ref{SecPrelim}, the actual scaleheights closely track the minimum of either the SG or NSG case when they are significantly different. On the other hand, when SG and NSG heights are comparable the true height is smaller than either of them. 

The MW thin disc thickness profile is indicated by the blue shaded area; its half-mass scaleheight is $\sim 100$ pc at the solar circle, and increases steadily with radius, reaching $\sim 450$ pc at $R=14$ kpc. In the outer regions these values match very well the expectation from our simple model. Towards the center the disc thins down, but more gradually than expected from our simple model. Note that this inward thinning (or outward ``flaring'') is naturally consistent with theoretical expectations for exponential discs in CDM halos, and not necessarily a result of secular evolutionary processes in the disc.

\subsection{Exponential discs in CDM halos: mass dependence}
\label{SecExpD}

We now consider the thickness of (gaseous) exponential discs as a function of disc and halo mass, assuming again that they are embedded in CDM halos. In our model, the problem is fully specified once the relations between (i) halo mass and disc mass, and between (ii) disc mass and size have been specified. Note that the halo concentration is not a free parameter, since it is a well understood function of halo mass once the cosmological parameters are fixed \citep[see, e.g.,][]{Ludlow2016}. 

The first relation is well constrained by the galaxy mass function. A simple parameterization is provided by ``abundance matching'' (AM) models, such as that of \citet{Behroozi2013}. The second relation is empirically well established and may be adequately described by the simple relation 
\begin{equation}
R_{h}=0.2\, r_s,
\label{EqRhRs}
\end{equation}
 where $R_{h}=1.678\,R_d$ is the galaxy half-mass radius and $r_s$ is the NFW scale radius of its surrounding AM halo \citep[see, e.g., Fig.1 of][]{Navarro2016}.

We use these relations to study the thickness of gaseous discs formed in $\Lambda$CDM halos. We begin by considering exponential discs whose radial scalelengths satisfy equation (\ref{EqRhRs}), and vary the disc mass, keeping the halo mass fixed. This is shown in Fig.~\ref{FigMdz50}, where each panel shows the half-mass scaleheight at $R_d$, as a function of the assumed disc mass, $M_{d}$, for three different values of the halo mass, $M_{200}$. Since the disc size is fixed in each panel, $M_d$ scales linearly with disc surface density, $\Sigma(R_d)$. 

The acceptable range of disc masses according to AM models is indicated by the shaded vertical band. There are three curves in each panel, corresponding to three different values of $\Gamma$: $1$, $4/3,$ and $2$, normalized as in equation (\ref{EqCsEoS}) so that the pressure of the systems at $\rho_{g} = \rho_{\rm eos}$ is $P/k_{B}=\rho_{\rm eos} T_{\rm eos}/m_{p}$, with $\rho_{\rm eos}/m_{p} = 0.1 \rm \ cm^{-3}$ and $T_{\rm eos} = 8000 \ K$. Curves change linewidth when the discs turn unstable; i.e., the Toomre parameter $Q(R_d)>Q_{\rm crit} \sim 1$ (equation \ref{EqToomre}) in the thick linewidth regime, and thin otherwise. 

As expected, low-mass discs are stable and non-self-gravitating but they all become unstable when the mass of the disc exceeds a critical, $\Gamma$-dependent value. This critical value exceeds the mass expected from AM arguments in low-mass halos; therefore, ``realistic'' galaxy discs (i.e., those that satisfy both abundance matching constraints and the empirical size-mass relation) forming in halos less massive than $\sim 10^{11}\, M_\odot$ are expected to be stable. Note that this is a conservative conclusion, since we are assuming that the whole mass of the galaxy is in gaseous form. If part of that mass was in stars this would decrease the gas density and reinforce the stability condition.  On the other hand, galaxy discs in halos of order or exceeding $\sim 10^{12}\, M_\odot$ are expected to be Toomre unstable, if assumed gaseous. 

These results have important implications for cosmological simulations that assume an EAGLE-like EoS, since it implies that stars will form more or less uniformly throughout the disc in low-mass systems, but in self-bound clumps in massive discs. We shall return to this issue in Sec.~\ref{SecSoftInst}.

Finally, we remark on how the $z_{50}$ dependence on mass varies as a function of $\Gamma$. For isothermal discs ($\Gamma=1$), $z_{50}$ is roughly constant when disc masses are low, and decreases monotonically with increasing mass (or, equivalently, with increasing surface density). 

For $\Gamma=4/3$ the sound speed is a function of density (equation~\ref{EqCsEoS}) and, therefore, the scaleheight increases with increasing mass  in the low-mass regime (denser discs are effectively ``hotter'') until reaching a maximum when the vertical contribution from the halo and the disc become comparable. At higher disc masses the disc thins down steadily because of the increased disc contribution, and eventually becomes unstable. 

For $\Gamma=2$ the behavior is qualitatively similar at low masses but the scaleheight asymptotically approaches a maximum as the disc becomes self-gravitating and, eventually, unstable. This is because self-gravitating disc scaleheights are proportional to $c_s^2/\Sigma$ and, therefore, to $\Sigma^{(\Gamma-2)/\Gamma}$ for a polytropic EoS.  For $\Gamma=2$, then, pressure and self-gravity balance each other out at a constant height, independent of disc mass.

We summarize these results in Fig.~\ref{FigMbarMh}, where we plot disc mass vs halo mass colored by half-mass scaleheight at $R_d$, assuming the $\Gamma=4/3$ fiducial EAGLE EoS and disc sizes given by equation (\ref{EqRhRs}). The top area (shaded in dark blue) indicates masses that exceed the total baryonic budget of the halo, $(\Omega_{\rm bar}/\Omega_{\rm M}) M_{200}$, and are excluded in CDM models. The thick dot-dashed line traces the abundance-matching relation. The thick solid line indicates the boundary between stable and unstable discs; i.e., where the condition $Q(R_d)=Q_{\rm crit}$ is satisfied.

\begin{figure*}
    \includegraphics[width=\textwidth]{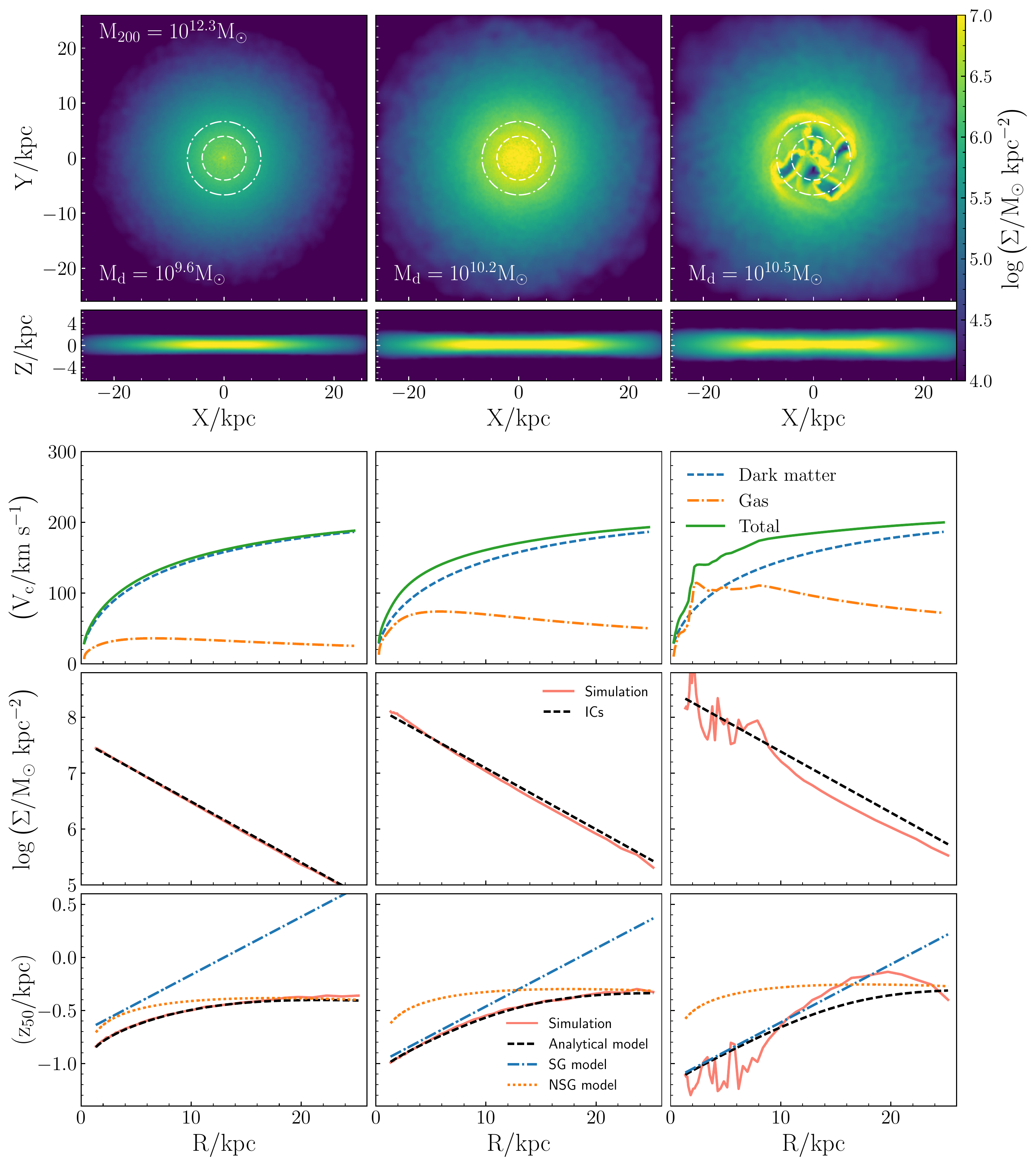}
    \caption{Three exponential polytropic discs evolved in a dark matter halo of mass $M_{200} \sim 2.0 \times 10^{12} \rm M_{\odot}$, for different gas masses (increasing from left to right, as indicated by the legends), but keeping the scale radius of the system fixed, $R_d=4$ kpc (inner dashed circle in top panels). For reference, the virial radius of the system is $r_{200}=260 \rm \ kpc$. These systems are highlighted with white squares in Fig.~\ref{FigMbarMh}. The left and middle columns correspond to stable discs evolved for $\sim 3 \rm \ Gyr$. The rightmost panel shows, in contrast, a disc that is Toomre unstable and breaks into clumps, evolved for $\sim 150 \rm \ Myr$. The bottom row of panels shows the circular velocity profile, $V_c(R)$; the surface density profile, $\Sigma(R)$; and the half-mass scaleheight, $z_{50}$, of the discs. Red solid curves show the measured profiles, whilst dark dashed lines show the analytic model. Note that the surface density profile evolves little from that imposed in initial conditions. The analytic model for the $z_{50}$ profile is calculated using the method described in Sec.~\ref{SecH}, using the initial surface density profile as input. Blue dot-dashed, and orange dotted lines show the solution for a self-gravitating and a non-self-gravitating disc, respectively.}
    \label{FigDiscs}
\end{figure*}

``Realistic'' discs (i.e., those matching the AM constraint) are generally stable at low masses and unstable at high masses. Their aspect ratios are also expected to be a strong function of disc mass, varying from  from $z_{50}/R_{d}$ of order $1$:$40$ for Milky Way-like discs to $1$:$5$ for $\sim 10^8 \, M_\odot$ discs. Note that this is a simple consequence of the scaling properties of gaseous discs and the assumed EoS, and {\it not} the consequence of limitations in  numerical resolution or other shortcomings of the hydrodynamical treatment.

In the next section we use these results to interpret the structure of galaxy discs simulated using some of the numerical techniques used in the  latest cosmological simulations. We focus, in particular, on how well these simulations reproduce the expected vertical structure of polytropic discs and the onset of radial instabilities, as well as the scaleheight differences between gaseous discs and the stellar discs they evolve into.

\section{Numerical simulations}
\label{SecSims}

Our tests follow the evolution of  polytropic gaseous discs embedded in the (rigid) potential of a dark matter halo. We consider two types of tests. In the first type the gas is not allowed to turn into stars and the simulation is followed for a prescribed number of orbital times or until much of the mass of the disc (if unstable) is in the form of distinct self-gravitating clumps. In the second type the gas denser than $\rho_{g}/m_{p} = 0.1\,\rm cm^{-3}$ is allowed to cool and turn into stars at the same Kennicutt-Schmidt rate adopted in the EAGLE suite of cosmological simulations \citep{Schaye2015,Crain2015}. For simplicity, we neglect the effects of enrichment and feedback from evolving stars into the surrounding gas.

\subsection{The code}
\label{SecCode}

We use the public version of the {\tt Gadget-2}\footnote{\tt wwwmpa.mpa-garching.mpg.de/gadget/} code, modified to include an NFW rigid spherical potential modelled after a CDM halo. This implies that the gas and stars in our simulations experience, in addition to their self gravity, a central gravitational acceleration equal to $V_{\rm dm}^2(<r)/r$, with
\begin{equation}
V_{\rm dm}^2(<r) = V_{200}^2\ {1\over x}\  {\ln(1+cx) - (cx)/(1+cx) \over \ln(1+c) - c/(1+c)},
\label{EqVNFW}
\end{equation}
where $x=r/r_{200}$ is the radius in units of the virial radius and the concentration, $c=r_{200}/r_s$, links the virial radius with the NFW scale radius, $r_s$. The concentration is a function of the virial mass, $M_{200}=V_{200}^2\, r_{200}/G$, which is well understood once the cosmological parameters are fixed \citep{Ludlow2016}.

We have also modified  {\tt Gadget-2} to include an effective equation of state that enforces a polytropic law relating gas pressure and density (equation \ref{EqEoS}). In practice,  this is done by replacing the entropy of the {\it i-th} gas particle by
\begin{equation}
A_i = P_{\rm eos} \left ( \displaystyle\frac{\rho_i}{\rho_{\rm eos}} \right )^\Gamma \rho_i^{-\gamma}
\end{equation}
where $\Gamma=4/3$ is the effective polytropic index, and $\gamma = 5/3$ is the ratio of specific heats. The parameters of our fiducial runs are summarized in Table~\ref{Table:fiducial_parameters}, and were chosen to match those implemented in the EAGLE simulation suite~\citep{Schaye2015, Crain2015}.

\subsection{Initialization}

Our aim is to build numerical realizations of centrifugally supported exponential discs embedded in dark matter halos, so we initialize the gas  according to the following density profile:
\begin{equation}
\rho_g(R,z) = \displaystyle\frac{M_{\rm d}}{4\pi R_{d}^2 z_d} \exp{(-R/R_d)} \sech^2{(z/z_{d,0})}.
\end{equation}
where the initial scale-height, $z_{d,0}$, is independent of $R$ and substantially larger than the expected equilibrium disc thickness, according to the results of Sec.~\ref{SecH}. Gas particles are given only azimuthal velocities that equal the circular velocity of the system at their corresponding cylindrical radii.

As expected, the initial disc collapses vertically and quickly settles into equilibrium. This settling leads to a minor rearrangement of the radial profile but the deviations from the desired exponential are mild and limited to the inner regions. Furthermore, we have explicitly verified that the results we report below are independent of the particular value of $z_{d,0}$ adopted.

The disc is allowed to evolve for several orbital times before it is analyzed, typically for $\sim 25\, t_{\rm orb} (R_{h})$, where $R_{h}$ is the disc's half-mass radius, and $t_{\rm orb}(R_{h})$ is the disc's orbital time at $R_{h}$ ($\sim 4$ Gyr for a disc scaled to the Milky Way). 

We carried out simulations covering a grid of values in the halo mass-disc mass plane. In particular, we considered models with halo masses in  the range $1.0 \times 10^{10} \le M_{200}/ M_{\odot} \le 2.0 \times 10^{12}$, and disc masses in the range $1.25 \times 10^{7} \le M_{\rm d}/M_{\odot} \le 1.0 \times 10^{11}$. The disc size is kept fixed for each halo mass, so that its half mass radius is simply $R_{h}=1.678 R_d=0.2\, r_s$. NFW scale radii are determined by the $M_{200}(c)$ relation corresponding to the Planck cosmological parameters \citep{Ludlow2016}. We report here the results for discs at a given halo mass of $M_{200} = 2.0 \times 10^{12}\,M_{\odot}$, varying the gas mass, as shown by the orange dots in Fig.~\ref{FigMbarMh}, but we have verified that the following results apply irrespective of halo mass.

\subsection{Numerical resolution}

Our fiducial runs use $2 \times 10^5$ equal-mass gas particles, and have spatial resolution adequate enough to resolve the characteristic disc scaleheight and the scale of radial instabilities, if present. The Jeans mass, $M_J\approx 5.5 \times 10^{8} \ M_{\odot}$, is constant for $\Gamma=4/3$ and resolved with at least a few hundred particles in the worst case.  In addition, we use a Plummer gravitational softening $\epsilon_g = 10 \rm \ pc$, which is much smaller than the characteristic height of the discs in all cases. Finally, we impose a minimum gas smoothing length, $h_{\rm min} = 10$ pc, and have explicitly checked that this has little or no effect on the thickness of simulated discs or on the scale of the onset of radial instabilities. We list the main numerical parameters of the fiducial runs in Table~\ref{Table:fiducial_parameters}.

\begin{table}
 \caption{Relevant parameters of our fiducial runs.}
 \label{Table:fiducial_parameters}
 \begin{tabular}{clllllll}
 \hline
 $\rho_{\rm eos}/m_{p}$ & $T_{\rm eos}$ & $\Gamma$ & $\epsilon_{g}$ & $h_{\rm min}$ & Npart \\
 \hline
 $0.1 \rm \ cm^{-3}$ & $8000 \ K$ & $4/3$ &$10 \rm \ pc$ & $10\,\rm pc$ & $2\times 10^5$ \\
 \end{tabular}
\end{table}

\section{Numerical results}
\label{SecNumRes}

Figure~\ref{FigDiscs} illustrates the final state of three of our simulated discs, all assumed to have the same size ($R_d=4.0$ kpc), and evolved in the same dark matter halo: $M_{200} \sim 2\times 10^{12} \ M_{\odot}$ and $c=8.0$. These disks are highlighted with open squares in Fig.~\ref{FigMbarMh}.

Discs in the left and center columns ($M_d=4\times 10^9\, M_\odot$ and $1.6\times 10^{10}\, M_\odot$, respectively) are shown after $\sim 4$ Gyr (which corresponds to $\sim 25$ orbits at the half-mass radius, indicated by the dot-dashed circle in the top panels) are ostensibly in equilibrium, and have actually evolved very little since their initial relaxation stage. The disc shown in the right column, on the other hand ($M_d=3.2\times 10^{10}\, M_\odot$), has clearly  become unstable after settling vertically and has broken into a number of distinct self-gravitating clumps. This disc is shown after just $\sim 150$ Myr of evolution.  

This behaviour is easily understood in terms of the analysis of Sec.~\ref{SecAnalRes}. The first two discs are stable according to the criterion laid out by equation (\ref{EqInstCond}); the last one is unstable. 

Note also that in all three cases the initial settling of the gas has led to only minor changes in the azimuthally-averaged surface density profile, shown in the second row of Fig.~\ref{FigDiscs} (dashed lines correspond to the initial configuration; the solid red line is the final profile).

Finally, the bottom row compares the half-mass scaleheight, $z_{50}$ (solid red line), with that expected from the analytic treatment. The dot-dashed and dotted lines indicate the results of assuming that the disc is SG or NSG, respectively. The black dashed line is the final result, computed using equation (\ref{EqZh}). The model predictions are computed assuming the initial gas profile, but should apply to the discs as shown, given how little  $\Sigma(R)$ is affected by the subsequent evolution. 

The agreement between simulations and the analytic solution is clearly quite good, even in the case where the disc has developed massive instabilities.  This is because the time scale for the disc to settle into a vertical hydrostatic equilibrium configuration is generally shorter that the time that takes for the clumps to develop.  Although Fig.~\ref{FigDiscs} refers to only one halo mass, we have verified that the same conclusions apply to all our runs, giving us confidence that the numerical treatment of our discs adequately captures the physical processes responsible for setting the thickness of gaseous discs.

\begin{figure}
\includegraphics[width=\columnwidth]{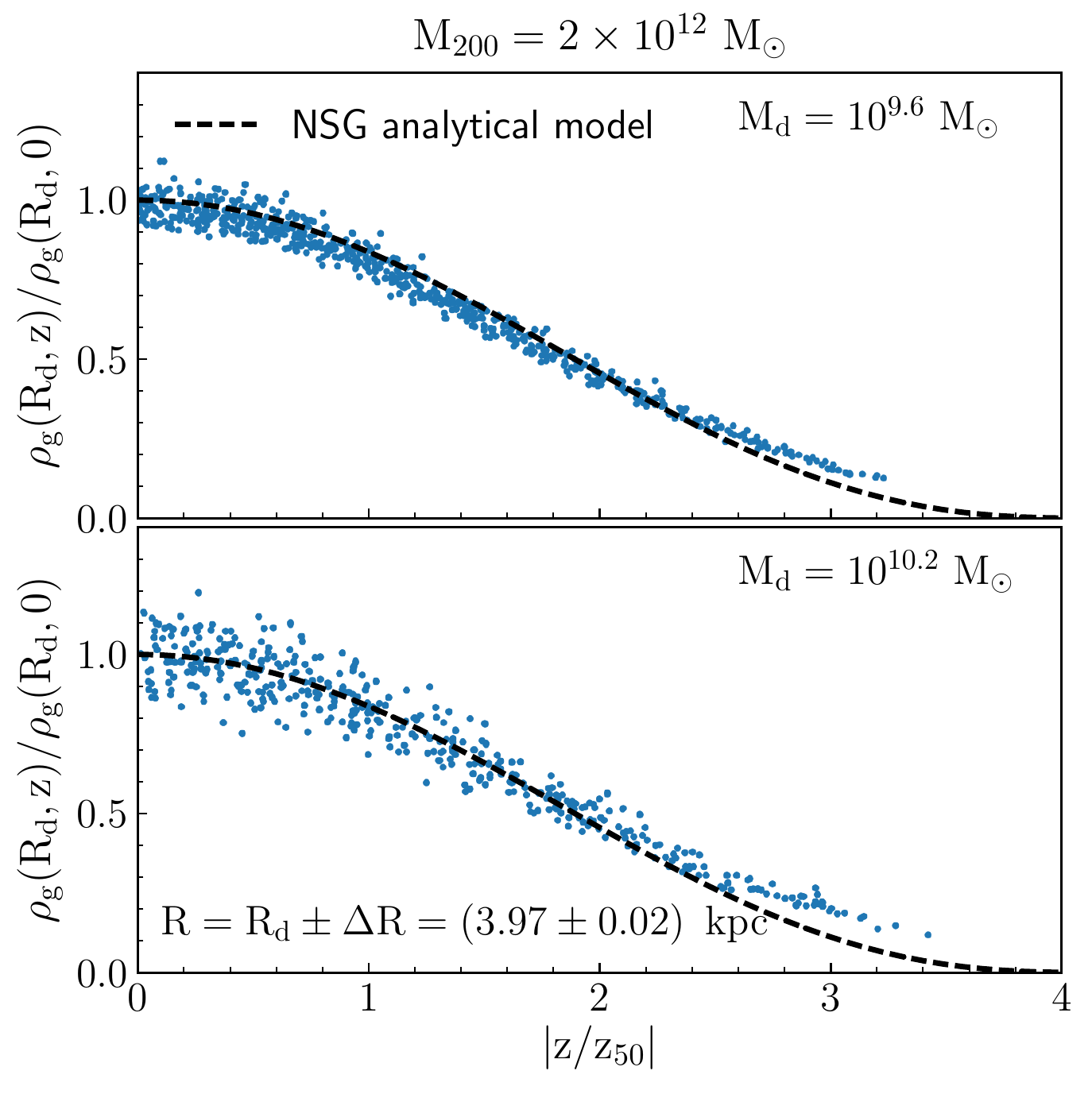}
\caption{Vertical density profile at the scale-radius, $R_d$, normalized to the density at the midplane of the disc, as a function of $z/z_{50}$, for two of the simulated discs shown in the left and middle columns of Fig.~\ref{FigDiscs}. The disc shown in the upper panel is non-self-gravitating, whilst the disc shown in the bottom panel is self-gravitating at $R_d$. Individual dots show the density of the gas particles within a small annulus centred at $R_{d}$. Black dashed-line shows the analytic density profile of a non-self-gravitating polytropic disc, given by equation (\ref{EqNSGRhoZ}). The simulations reproduce the analytic expectation very well. This also shows that the {\it shape} of the vertical density profile can be approximated by the equation (\ref{EqNSGRhoZ}), even when the system is self-gravitating.}
\label{FigRhoZSim}
\end{figure}

\begin{figure}
\includegraphics[width=\columnwidth]{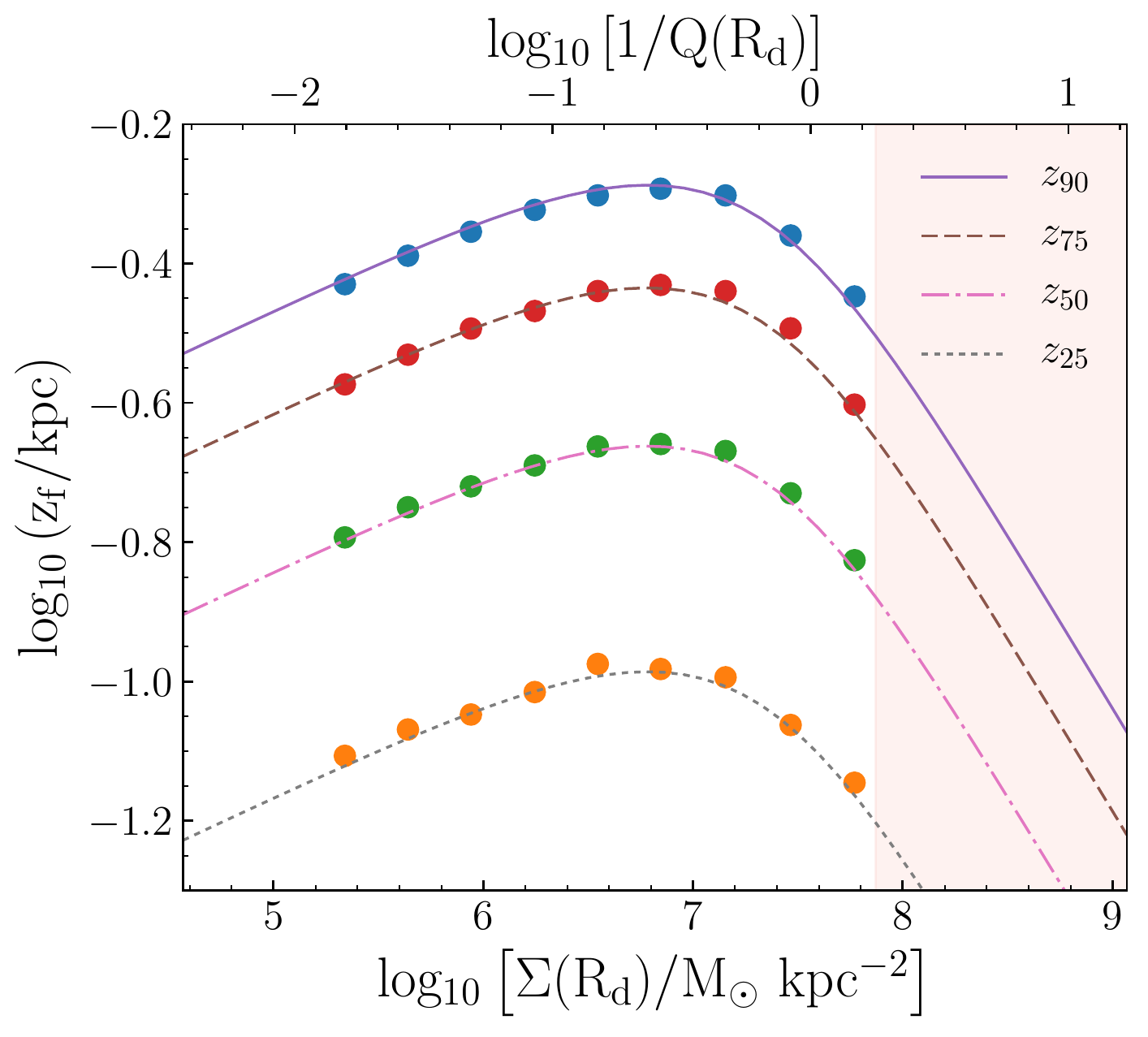}
\caption{The $z$-coordinate that contains a given fraction of the disc's column mass, $z_f$, as a function of surface density at $R_d$ for simulated discs evolved in the same dark matter halo of $M_{200} = 2\times 10^{12} \ M_{\odot}$. All discs have the same scale radius $R_d=4$ kpc, but varying masses: these correspond to all orange circles in Fig.~\ref{FigMbarMh} below (and including) the instability line. The shaded region indicates surface densities above which discs are expected to be unstable; circles are the results of the simulations; lines are the analytic predictions from Sec.~\ref{SecAnalRes}. The scale on top indicates the value of $1/Q(R_d)$ for each disk (from equation \ref{EqToomre}), which is a measure of the critical surface density above which discs become unstable.}
\label{FigZf}
\end{figure}

\subsection{Vertical density profiles}
\label{SecRhoz}

One interesting result from the solutions presented in Sec.~\ref{SecAnalRes} is that the {\it shape} of the vertical density profiles of polytropic discs (scaled to the midplane density and to the characteristic scaleheight) is a well specified function of $\Gamma$ and, to first order, independent of whether the disc is self-gravitating or not. We show this in Fig.~\ref{FigRhoZSim}, where we plot the vertical density profiles of the two galaxies in Fig.~\ref{FigDiscs} that do not develop strong radial instabilities and for which the vertical profiles can be meaningfully measured. 

The profiles are shown  at $R_{d} \sim 4$ kpc, and normalized to the midplane density and the half-mass scaleheight. At this radius, the scaleheight of one galaxy is determined by the NSG solution but the other is much closer to the SG regime (see bottom panels of Fig.~\ref{FigDiscs}). Regardless, their vertical profile {\it shapes} are quite similar, and may be well approximated by the NSG vertical density law (equation~\ref{EqNSGRhoZ} for $\Gamma=4/3$), shown by the thick dashed line in each panel of Fig.~\ref{FigRhoZSim}.  

The simulated profile shapes are clearly very similar in both cases, confirming our earlier conclusion that, appropriately scaled, the vertical dependence of the density is similar for SG and NSG discs.

An interesting corollary of adopting a $\Gamma=4/3$ polytropic EoS is that the sound speed decreases with decreasing density. As a result, and as shown in Fig.~\ref{FigMdz50}, the scaleheight of low-mass (NSG) discs should increase with $M_d$, reach a maximum, and then decrease as the self gravity of the disc becomes gradually more important. 

We show this in Fig.~\ref{FigZf}, where we plot the $z_f$ coordinate that contains different fractions of the disc column mass, measured at $R_d$, for runs that vary the disc mass, keeping the same halo and disc radius as in Fig.~\ref{FigDiscs}. The ``scaleheights'' $z_f$ are shown as a function of the surface density, $\Sigma(R_d)$, which is directly proportional to $M_d$, since the disc radial scalelength is kept fixed. As expected, low mass discs get thicker with increasing $\Sigma$, reach a maximum and then start to decrease at higher masses, before eventually becoming unstable.

The excellent agreement between the numerical results (coloured circles) and the analytic predictions (lines) give us confidence that the vertical structure of polytropic discs can be simulated accurately by current hydrodynamical techniques, at least when the spatial and mass resolution is adequate enough to resolve the Jeans mass and the characteristic thickness of the disc. Systematic deviations may occur, however, when simulated discs are evolved with inadequate resolution, such as, for example, when the gravitational softening is comparable to the disc thickness. We explore this next.

\begin{figure}
\includegraphics[width=\columnwidth]{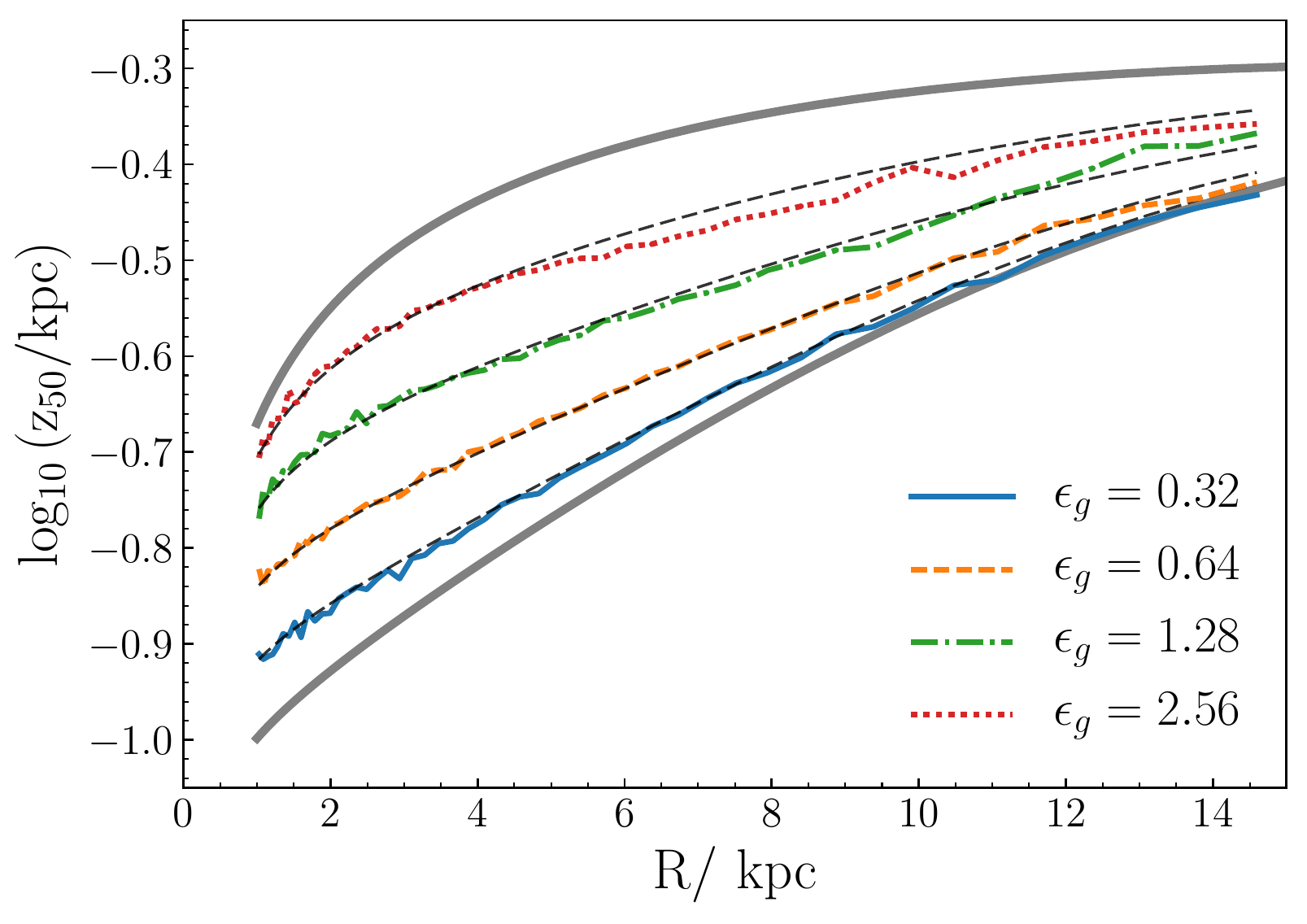}
\caption{Half-mass scaleheight, as function of radius, for the gas disc shown in the middle panel of Fig.~\ref{FigDiscs}, simulated with various values of the gravitational softening, $\epsilon_{g}$, in kiloparsecs, as indicated in the legend. Thin dashed curves close to the coloured curves show the predictions from equation (\ref{Eq:General_height_softened}). Thick top and bottom curves indicate the analytic NSG and SG  half-mass heights, respectively.} 
\label{FigZEps}
\end{figure}

\subsection{Disc thickness and gravitational softening}
\label{SecSoftening}

The gravitational softening, $\epsilon_g$, introduces a physical length scale expected to induce deviations from the analytic solutions in discs whose thickness is comparable to $\epsilon_g$. This will only affect self-gravitating discs, since the vertical self-gravity of the disc is, by definition, negligible in the non-self-gravitating case. Note that this implies that the role of $\epsilon_g$ should be negligible in cosmological simulations of the formation of low-mass discs that adopt an EAGLE-like EoS, as discussed in Sec.~\ref{SecExpD}: ``realistic'' discs less massive than $\sim 2.5\times 10^{11} M_\odot$ are expected to be non-self-gravitating.

\begin{figure}
\includegraphics[width=\columnwidth]{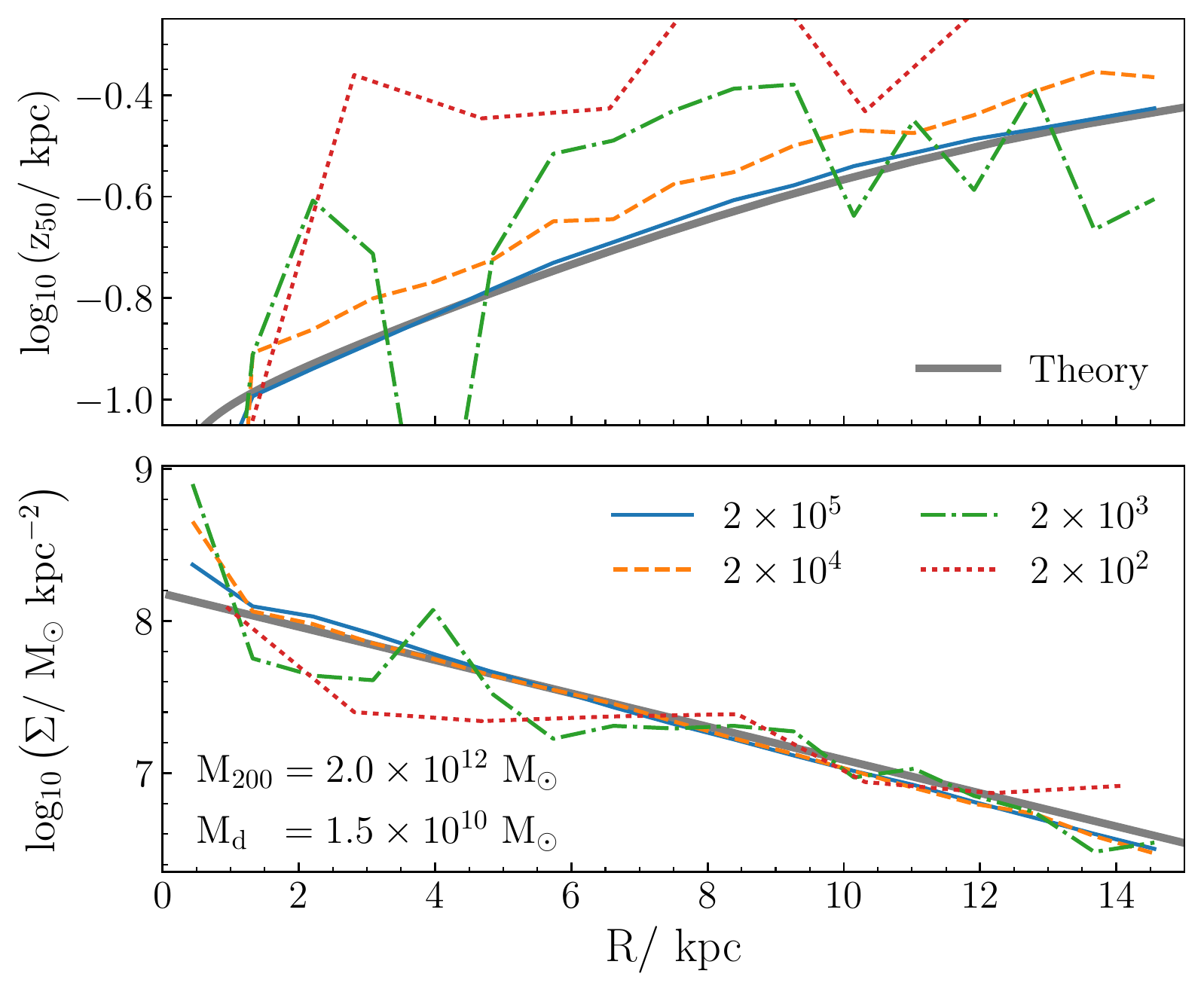}
\caption{Half-mass scaleheight (top) and surface density (bottom) as a function of radius for the self-gravitating disc shown in the middle panels of Fig.~\ref{FigDiscs}. The same disc is simulated varying the number of particles used to resolve the system, from $2\times 10^{5}$ to $200$ gas particles, as indicated by the legends in the bottom panel. Grey thick solid lines show the theoretical expectation for the input surface density profile.}
\label{FigThickRes}
\end{figure}

On the other hand, the gravitational softening may impact SG discs in two ways: (i) it may result in artificially thicker discs by reducing their vertical gravity, and (ii) it may hinder the development of disc instabilities by limiting the binding energy of the self-bound clumps that result. As discussed in Sec.~\ref{SecPrelim}, the length scale of radial instabilities is generally much larger than the scaleheight of the discs, so the condition $\epsilon_g << H_{\rm SG}$ should be enough to ensure good numerical convergence. This implies a minimum gravitational force resolution of $\sim 100$ pc to resolve the thin disc of the Milky Way at the solar circle and beyond. 

The above considerations may be tested by running a controlled series of simulations where the gravitational softening of a nominally SG disc (at $R_d$) is monotonically decreased until convergence is achieved. In this case, the actual thickness of a simulated disc is expected to vary from the non-self gravitating solution for large softening (because in that case the vertical gravity of the disk is negligible) until converging to the self-gravitating case for small enough softening.

We choose for this example the system depicted in the middle column of Fig~\ref{FigDiscs}, and show in Fig.~\ref{FigZEps} the radial dependence of the half-mass scaleheight for different values of $\epsilon_g$. The two thick lines at top and bottom illustrate the expected solution for the NSG case and the ``true'' ($\epsilon_{g} \ll H_{\rm SG}$) case, respectively. 

The simulation results clearly reproduce the expected trend, and also suggest a simple empirical formula to describe quantitatively the spurious thickening induced by the softening\footnote{We refer the reader to App.~\ref{App:general_scaleheight} for a derivation of equation (\ref{Eq:harmonic_mean_2_soft_paper}).}:
\begin{equation}
\label{Eq:harmonic_mean_2_soft_paper}
H^2 = \left ( \displaystyle\frac{1}{H_{NSG}^2}+\displaystyle\frac{\xi(\epsilon_g)}{2 H_{NSG}H_{SG}}+\displaystyle\frac{\xi(\epsilon_g)}{H_{SG}^2} \right )^{-1}
\end{equation}
where $\xi(\epsilon)$ is a correction function of the form:
\begin{equation}
 \xi(\epsilon_{g}) = \displaystyle\frac{1}{1+\left (\epsilon_{g}/H_{SG} \right)^{\nu}}.
\label{Eq:Correction_function_paper}
\end{equation}

The dashed lines in Fig.~\ref{FigZEps} are simply computed using the corrected (softening-dependent) formula (equation \ref{Eq:harmonic_mean_2_soft_paper}) with $\nu=1.4$.

This result also provides guidance for the choice of gravitational softening in cosmological simulations that adopt a polytropic EoS, such as EAGLE. Realistic disc galaxies (i.e., that satisfy AM and size constraints) have surface densities that increase with mass roughly as $\Sigma\propto M_d^{1/3}$. Since the scaleheight of SG polytropic discs scales, for $\Gamma=4/3$, as $z_{50}\propto \Sigma^{(\Gamma-2)/\Gamma}\propto M^{-1/6}$ (Sec.~\ref{SecExpD}), then {\it self-gravitating} EAGLE discs are expected to have roughly constant thickness, at least if in gaseous form. Therefore, in practice, choosing $\epsilon_g$ so that the vertical scaleheight of the least massive disc expected to be unstable is properly resolved ensures that all discs are adequately resolved, regardless of mass. For the fiducial EAGLE EoS, this mass is of order $M_d \sim 3\times 10^9\, M_\odot$ (Fig.~\ref{FigMbarMh}), which has a half-mass scaleheight of $z_{50}(R_d)\sim 125$ pc. Choosing $\epsilon_g$ somewhat smaller than this value seems safe.  Much smaller softenings would unduly increase computing time and the likelihood of incurring integration errors and are unnecessary, at least for this EoS.

\subsection{Disc thickness and mass resolution}

We can use the same disc as in the previous subsection to assess the sensitivity of our results to the mass resolution (i.e., the number of particles) of the simulation. For reference, we recall that the Jeans mass for the EAGLE EoS is a constant, $M_J=5.5 \times 10^8\, M_\odot$. The reference disc has $M_d=1.5\times 10^{10}\, M_\odot$, which implies that our simulations resolve the Jeans mass in this system with at least a few hundred particles. We show in Fig.~\ref{FigThickRes} the half-mass scaleheight of the disc, as well as its surface density, in runs with varying particle numbers, from an extremely poorly resolved system with $200$ particles to our fiducial runs with $200,000$ particles.

Interestingly, except perhaps for the inevitable noise associated with poor particle sampling, both the vertical height profile and the surface density profiles are roughly in agreement with each other and with the analytic expectations. This suggests that, to first order, the discreteness effects associated with a finite number of gas particles are not the main systematic effect that limits the reliability and applicability of numerical simulations of disc galaxy formation, beyond the noise introduced by small number statistics.

\begin{figure}
\includegraphics[width=\columnwidth]{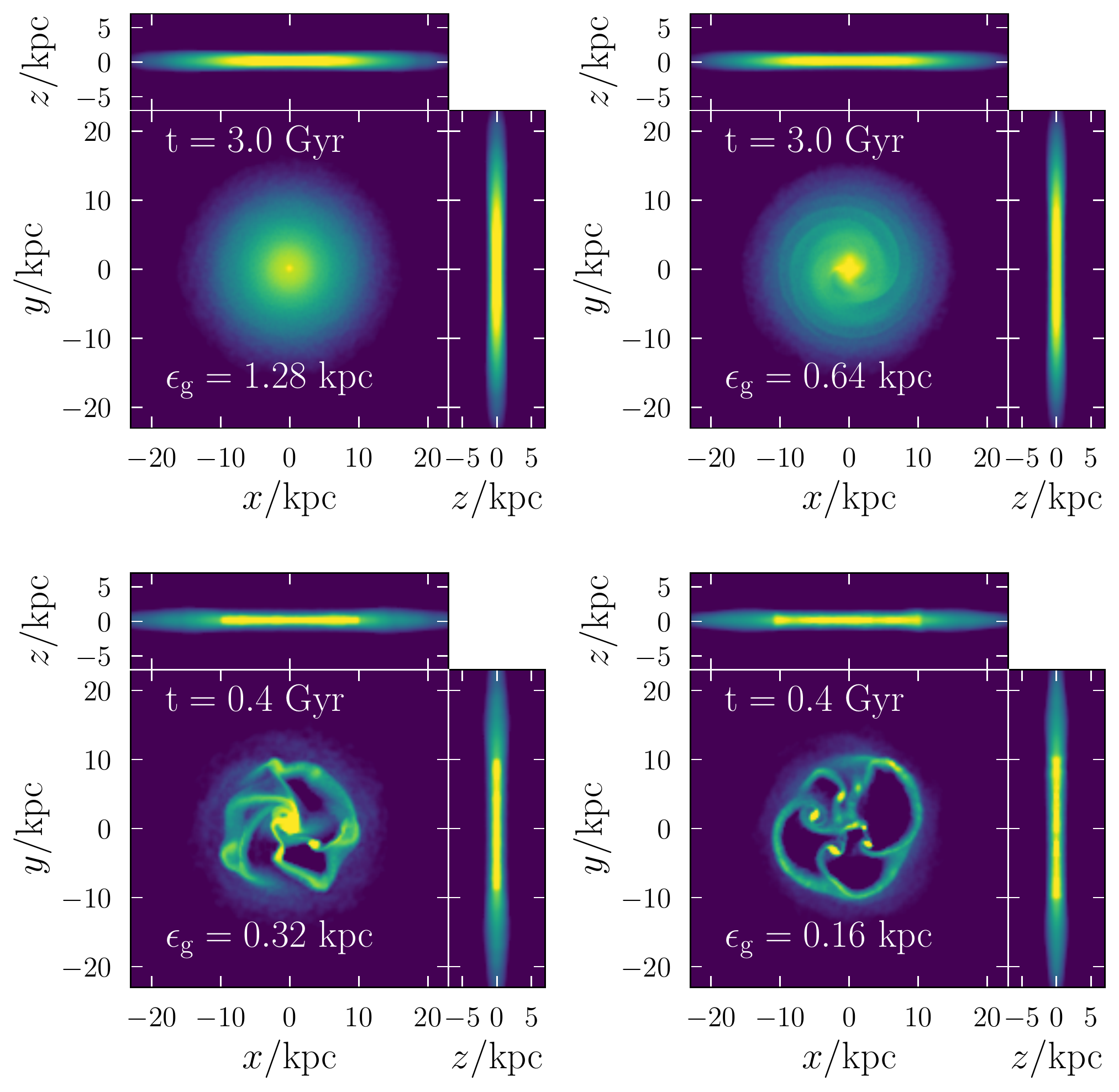}
\caption{Configuration of the gas in the nominally unstable gas disc shown in the right-hand column of Fig.~\ref{FigDiscs}. Runs differ only in the value of the gravitational softening, as given in the legends. For reference, the expected half-mass scaleheight of this disk at $R_d$ is of order $100$ pc.  The top row shows that the disc may be artificially stabilized when using large values of the gravitational softening, $\epsilon_{g}$. The bottom row shows that instabilities develop when $\epsilon_{g}<\epsilon_{\rm crit} \sim 0.39 \rm \ kpc$  (equation~\ref{EqEpsCrit}). This illustrates that instabilities develop when the critical radial wavelength, $\lambda_{\rm crit}$, is well resolved, even if the vertical scaleheight is not well resolved. Runs that become unstable are shown just after the clumps start to dominate the dynamics of the disc.}
\label{FigZEps2}
\end{figure}

\begin{figure}
\includegraphics[width=\columnwidth]{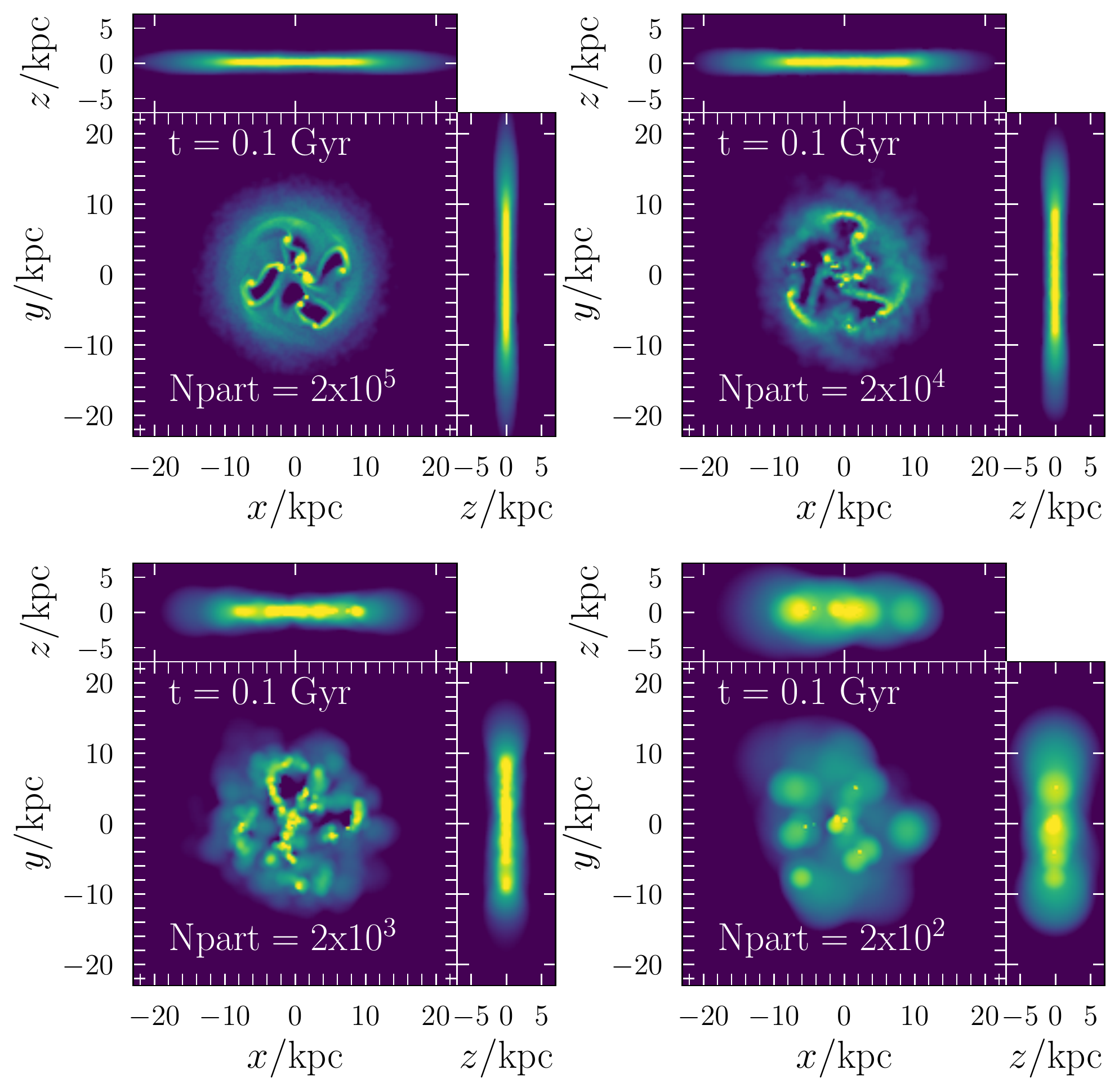}
\caption{As Fig.~\ref{FigZEps2}, but for $\epsilon_g=10$ pc and varying the number of particles used in the simulation (see legends in each panel). Note that for $\epsilon_g<\epsilon_{\rm crit}$ the same instabilities develop in the disc, regardless of the number of particles used. The gas particle mass is, in each run, $m_{\rm gas} = 1.5 \times 10^{5}$, $1.5 \times 10^{6}$, $1.5 \times 10^{7}$ and $1.5 \times 10^{8} \ M_{\odot}$, respectively.  For reference, the Jeans mass is the same in all cases, $M_J\sim 5.5 \times 10^{8} \ M_{\odot}$, and is resolved with fewer than 5 particles in the most poorly resolved run.}
\label{FigInstRes}
\end{figure}

\begin{figure*}
\includegraphics[width=\textwidth]{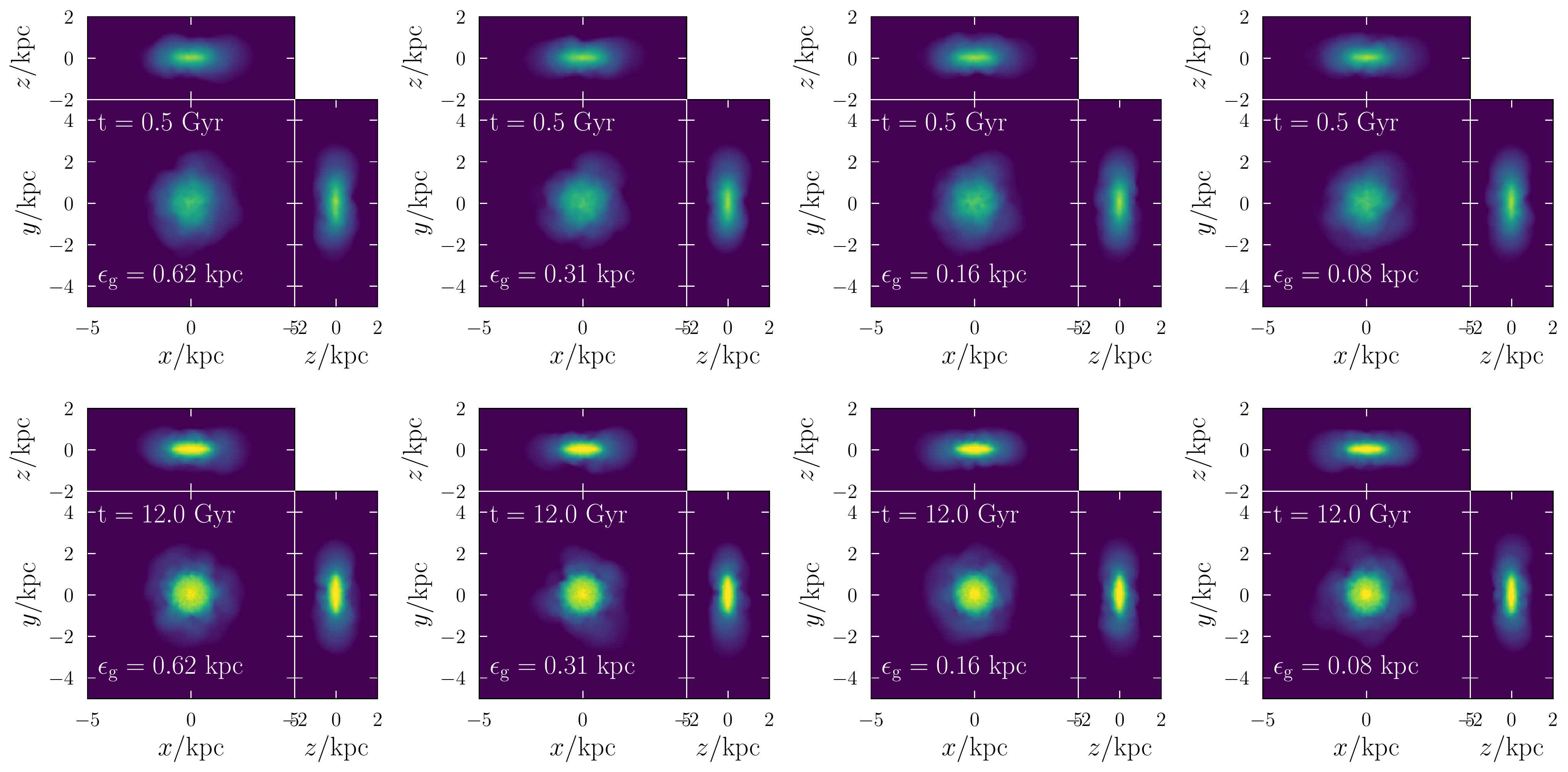}
\caption{Spatial configuration of the {\it stellar} disc formed from a parent gas disc that was nominally stable. The parent system is indicated by the leftmost starred symbol in Fig.~\ref{FigMbarMh}: $M_{200} = 1.0\times10^{10} \ M_{\odot}$, $M_{d} = 1.0 \times 10^{8}\,M_{\odot}$, $R_{d} \sim 0.9 \rm kpc$. Each column shows the same system at two different times, for simulations with gravitational softening, $\epsilon_g$. Varying the gravitational softening for more than a decade has little effect on the final configuration of the gas since the disk is nominally stable.}
\label{FigStellarD1}
\end{figure*}

\subsection{Disc instabilities and gravitational softening}
\label{SecSoftInst}

As anticipated in our discussion of Sec.~\ref{SecPrelim}, disc instabilities introduce another characteristic length scale in the problem: $\lambda_{\rm crit}=4\pi^2G\Sigma/\kappa^2$, the critical wavelength that arises in linear stability analysis of differentially rotating discs. Simulations with numerical resolution unable to resolve this scale are likely to miss the onset of radial instabilities and artificially stabilize the disc. 

The critical wavelength is typically larger than the disc thickness, so simulations that properly resolve the scaleheight of unstable discs will also capture the onset of radial instabilities. On the other hand, discs artificially thickened by limited resolution might see their instabilities suppressed once the gravitational softening becomes larger or comparable to $\lambda_{\rm crit}$.

We explore this in Figure~\ref{FigZEps2}, where we show the evolution of  a nominally unstable disc (the same one as in the right-hand column of Fig.~\ref{FigMbarMh}) for different values of $\epsilon_g$. The top panels show the final structure of discs that remain stable over many rotation periods, whereas the bottom panels show discs that  can only be followed for a few rotations because they quickly go unstable after settling in vertical equilibrium.

The {\it only} difference between the top/bottom rows is the value of the gravitational softening, confirming that gaseous discs may be artificially stabilized when the softening parameter exceeds a certain ``critical'' value. In the example of Figure~\ref{FigZEps2}  this ``critical'' value is of order $\epsilon_{\rm crit}\approx 0.39$ kpc\footnote{$\epsilon_{\rm crit}$ is calculated using quantities measured at $R_d \sim 4 \rm \ kpc$, where the disc is expected to be unstable. In particular, we use $\kappa(R_d) \approx 2V_{c}(R_d)/R(R_d) \sim 90 \rm \ km/s/kpc$ and $\Sigma(R_d) \sim 10^{8} \ M_{\odot} / \rm kpc^2$ (see Fig.~\ref{FigDiscs}).} This is actually larger than the expected half-mass scaleheight of the disc, $z_{50}(R_d)\approx 150$ pc, as shown in Fig.~\ref{FigDiscs}. In other words, simulations that resolve the critical instability wavelength go unstable even if they overestimate the thickness of the disc. We conclude that, as expected, it is easier to resolve radial instabilities in a nominally unstable gaseous disc than its scaleheight.

Although we focus on a single example in Fig.~\ref{FigZEps},  we have verified that the simple relation,
\begin{equation}
\epsilon_{\rm crit}={1\over 6}\, \lambda_{\rm crit}={2\pi^2 \over 3}{G\Sigma\over \kappa^2}
\label{EqEpsCrit}
\end{equation}
describes well the transition from stable to unstable discs in all our simulations.  The $(1/6)$ constant in the definition of $\epsilon_{\rm crit}$ may be understood by noting that (i) the ``most unstable'' wavelength is actually $(1/2)\lambda_{\rm crit}$ \citep[Sec. 6.2.3 of][]{Binney2008}, and (ii) that we are quoting ``Plummer-equivalent'' values for the softening. In these units, pairwise Newtonian gravity is recovered at distances $\sim 3\epsilon_g$. The two factors readily explain the $(1/6)$ constant relating $\epsilon_{\rm crit}$ and $\lambda_{\rm crit}$. We refer the reader to App. ~\ref{App:Instabilities} for a derivation of Equation (\ref{EqEpsCrit}).

\subsection{Disc instabilities and mass resolution}
\label{SecInstNpart}

Fig.~\ref{FigInstRes} shows the result of runs of the same system as in the previous subsection, but fixing the gravitational softening to $\epsilon_g=10$ pc, and varying the number of particles. This is a similar exercise to that illustrated in Fig.~\ref{FigThickRes}, where we showed that the expected disc thickness is well reproduced even with as few particles as a few hundred. As  Fig.~\ref{FigInstRes} makes clear, poor mass resolution does not alter the fundamental unstable structure of the disc, provided that $\epsilon_g<\epsilon_{\rm crit}$, which ensures that the radial instability is not artificially damped out by the softened gravity. We conclude that the number of particles does not impose critical restrictions on the structure of a gaseous disc, beyond those associated with the noise resulting from the finite particle number.

\begin{figure*}
\includegraphics[width=\textwidth]{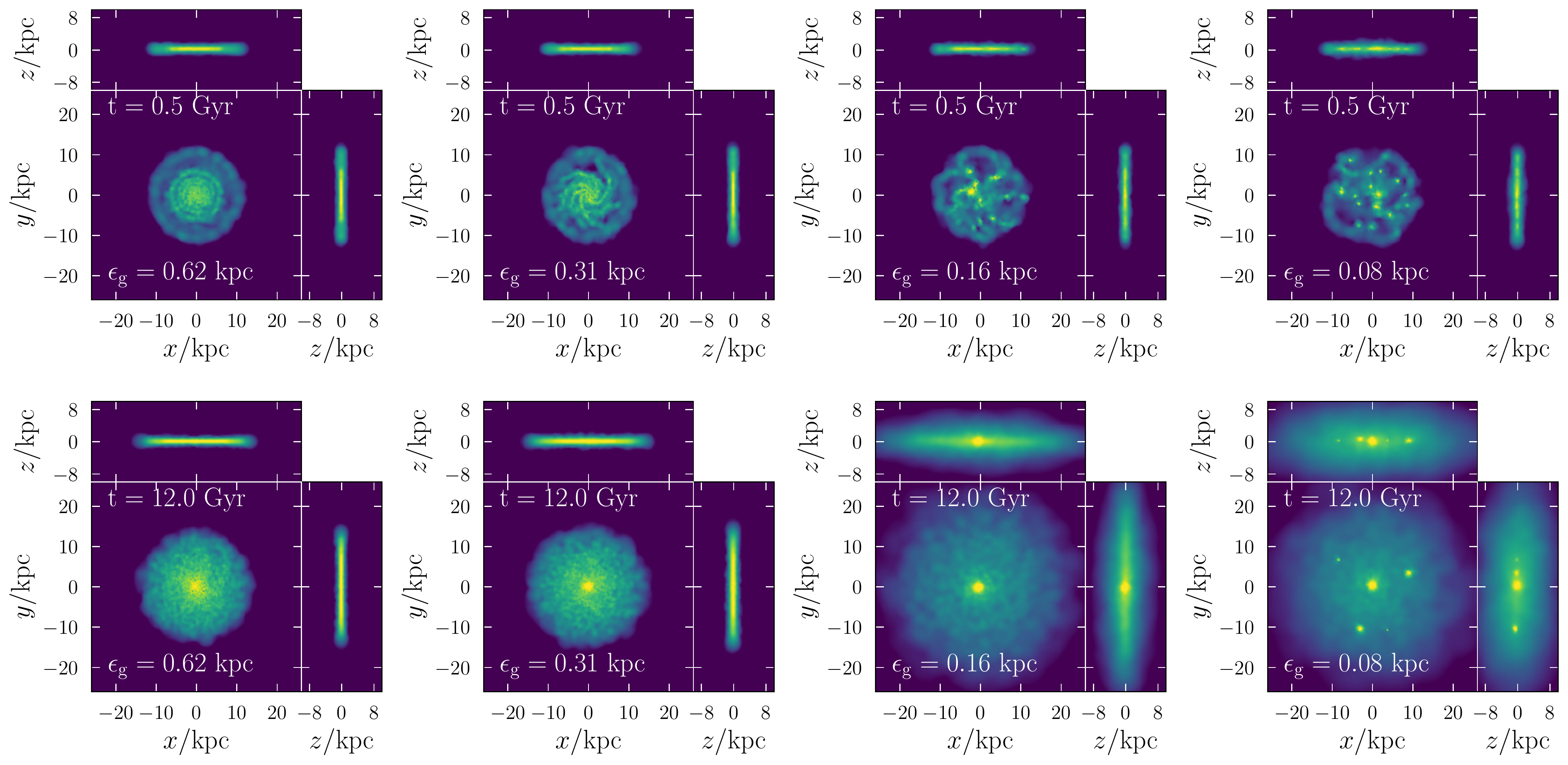}
\caption{Spatial configuration of the {\it stellar} disc formed from a parent gas disc that was nominally unstable. The parent system is indicated by the rightmost middle starred symbol in Fig.~\ref{FigMbarMh}: $M_{200} = 1.0\times10^{12} \ M_{\odot}$, $M_{d} = 2 \times 10^{10}\,M_{\odot}$, $R_{d} \sim 4 \rm kpc$. Each column shows the same system at two different time, simulated varying the gravitational softening, $\epsilon_g$. Note that when $\epsilon_g<\epsilon_{\rm crit}=0.5$ kpc (see blue arrow in bottom panel of Fig.~\ref{FigStellarDSoft}), instabilities may develop and the system breaks into self-bound clumps that subsequently disperse by merging, leading to considerable vertical thickening.} 
\label{FigStellarD2}
\end{figure*}

\subsection{The stellar descendants of polytropic gaseous discs}
\label{SecStellarD}

Stellar discs inherit the properties of the gas at the time of formation, but they evolve differently since, once formed, they are not subject to hydrodynamical forces. As such, stellar discs are subject to different kinds of instabilities and may evolve differently over time than their gaseous components. This is a complex topic beyond the scope of this paper, but two questions are of interest and we briefly consider them here: (i) how does the vertical scaleheight of a stellar disc reflect that of its parent gas?, and (ii) what are the effects of radial instabilities in the gas on the structure of the descendent stellar disc? 

We address these questions using a simplified approach, which relies on a parallel series of simulations identical to the ones discussed so far, but where we allow the gas that exceeds a density threshold, $\rho_{\rm thr}/m_{p} = 0.1\,\rm cm^{-3}$, to turn into stars at the same Kennicutt-Schmidt rates adopted in the EAGLE suite of cosmological simulations ~\citep{Schaye2015, Crain2015}\footnote{This is done by using a simplified version of the code used to run the EAGLE simulations.}. To simplify the problem further we neglect any feedback from formed stars. In other words, the gaseous discs are allowed to transform gradually into stellar systems until they essentially run out of gas, and we focus our analysis on their vertical equilibrium structure.

We begin by analyzing the vertical structure of a ``realistic'' disc in a low-mass halo; i.e., with gas mass and size that approximately match observational constraints. In particular, we choose for this illustration $M_{200}=3 \times 10^{10}\, M_\odot$, $M_d=10^{8}\, M_\odot$,  and $R_d=0.9$ kpc (leftmost starred symbol in Fig.~\ref{FigMbarMh}).  We expect such discs to be non-self-gravitating and, therefore, stable. In addition, since its scaleheight is determined by the halo and not by the disc's own gravity, the role of the gravitational softening in the vertical structure is negligible, simplifying the interpretation.

Each column of Fig.~\ref{FigStellarD1} corresponds to different runs of the same system, varying systematically the gravitational softening, from $\epsilon_g=0.62$ kpc (leftmost) to $0.08$ kpc (rightmost). For reference, the gas half-mass scaleheight of this system is expected to be $z_{50}(R_d) \sim 170$ pc. Each run is shown at two different times; the first one (top row) is shortly after the disc has settled vertically and when only about $5\%$ of the gas has been transformed into stars. The bottom row shows the system after several dozen rotation periods, long after star formation has effectively ceased. 

\begin{figure}
\includegraphics[width=\columnwidth]{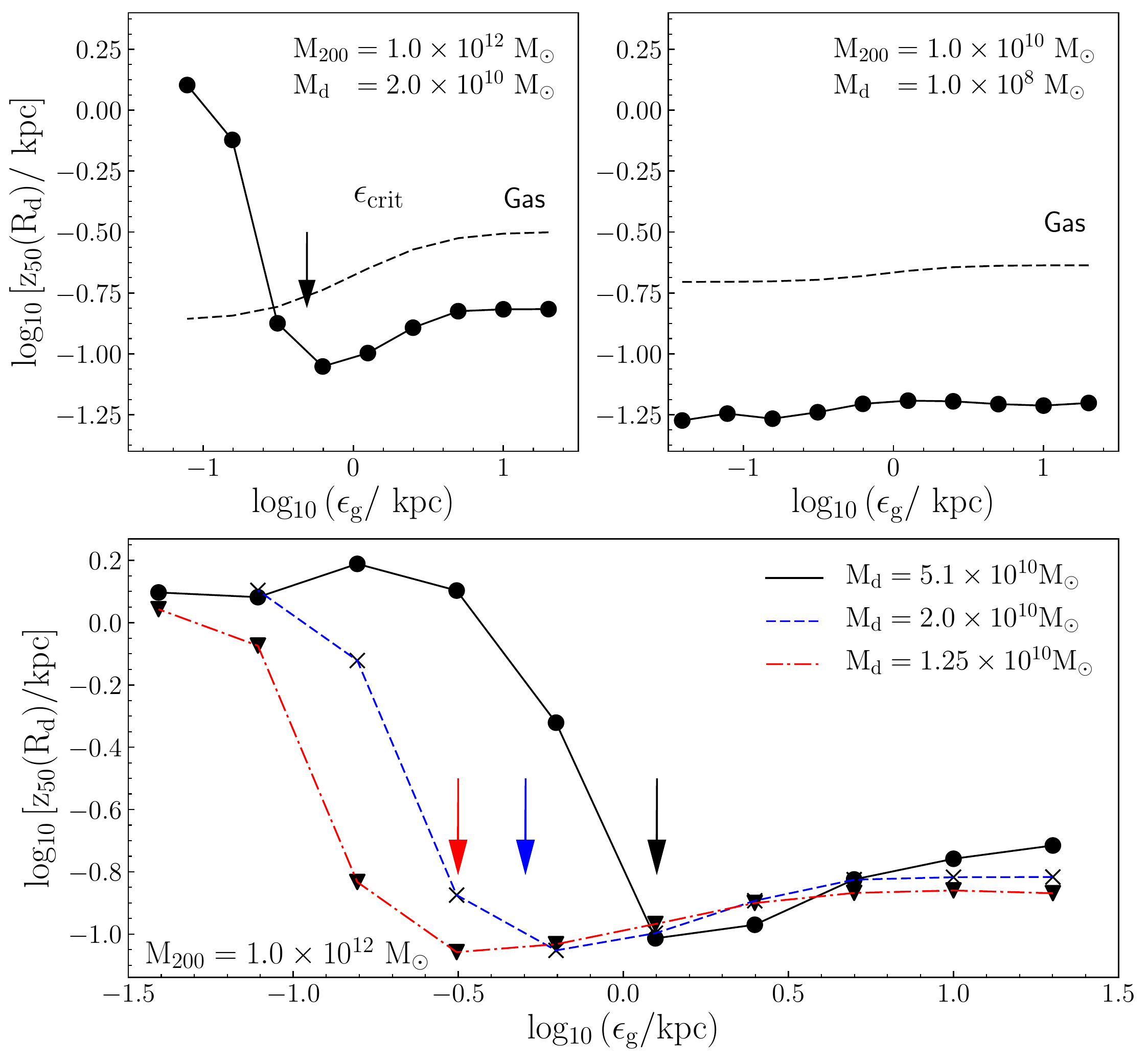}
\caption{{\it Top:} Half-mass scaleheight as a function of gravitational softening for two {\it stellar} discs formed out of a nominally unstable parent disc (left panel) or a stable one (right panel). Stellar discs formed out of stable gas disks  in vertical equilibrium are considerably thinner than the gaseous progenitor, by more than a factor of $\sim 2$, and independent of the value of $\epsilon_g$, even when this is varied by {\it three decades}. The same is true of nominally unstable disks artificially stabilized by large softenings. For $\epsilon_g<\epsilon_{\rm crit}$, though, the opposite is true: stellar discs are much thicker than their gaseous counterparts, an effect traced to the dispersal of self-bound clumps that assemble in these systems. {\it Bottom:} Same as top-left, but for three nominally unstable discs with different values of $\epsilon_{\rm crit}$, shown with vertical arrows. The scale radius is kept equal ($R_d = 4\,\rm kpc$). Stellar discs with gravitational resolution enough to resolve the size of unstable clumps are much thicker than their parent gas discs.}
\label{FigStellarDSoft}
\end{figure}

The choice of gravitational softening is clearly of little importance for the system shown in Fig.~\ref{FigStellarD1}, and the stellar discs at late time are practically indistinguishable amongst them. The disc thickness remains roughly constant, even after varying $\epsilon_g$ by  a large factor. 
We show this quantitatively in the top-right panel of Fig.~\ref{FigStellarDSoft}, where we plot $z_{50}(R_d)$ as a function of $\epsilon_g$ after $12$ Gyr of evolution. The thickness of the stellar disc is essentially independent of $\epsilon_g$, even as this is varied by three {\it decades}. This confirms that the gravitational softening plays essentially no role in the scaleheight of a non-self-gravitating disc.

It is also apparent from the same top-right panel of Fig.~\ref{FigStellarDSoft} that the stellar disc is actually {\it thinner} than its gaseous progenitor. There are two main reasons for this: one is that stars form faster in high-density regions and, consequently, form preferentially close to the disc midplane. The second is that stars are born out of gas in hydrostatic equilibrium and, therefore, at rest vertically; the stellar disc must therefore thin down by roughly a factor of two before virializing.

The situation is quite different when a more massive, nominally unstable disc is allowed to form and turn into stars. We show this in Fig.~\ref{FigStellarD2}, which is analogous to Fig.~\ref{FigStellarD1} but for $M_{200}=1.0\times 10^{12} \, M_\odot$, $M_d=2.0 \times 10^{10} \, M_\odot$,  and $R_d = 4 $ kpc. As discussed in the previous subsection, large values of  the gravitational softening may (artificially) stabilize the disc, leading to results qualitatively similar to those obtained for the low-mass system discussed above. This is indeed the case for the first two columns in Fig.~\ref{FigStellarD2} where the softening exceeds the ``critical'' value given by equation (\ref{EqEpsCrit}), which, at $R_d$, is $\epsilon_{\rm crit}(R_d)\approx 500 $ pc for this system.

For softenings smaller than the critical value, the growth of radial instabilities is no longer impeded and the disc quickly breaks up into a number of self-bound clumps that then turn into stars. These massive clumps orbit within the disc, colliding frequently, merging,  and eventually dispersing to leave behind a kinematically hot, much thicker stellar disc. 

This is seen qualitatively in the two columns on the right of Fig.~\ref{FigStellarD2}, and quantitatively in the top-left panel of Fig.~\ref{FigStellarDSoft}. For all runs with $\epsilon_g<\epsilon_{\rm crit}$, the stellar disc is nearly an order of magnitude thicker than otherwise. The sharp transition between these two regimes is indicative of the onset of an instability when the softening is small enough to resolve the formation of tightly bound clumps of stars in the disc.

If our interpretation is correct then the softening required to resolve radial instabilities should scale like $\Sigma(R_d)$ (equation~\ref{EqEpsCrit}) when all other parameters are kept equal. We test this by repeating the simulation series depicted in  Fig.~\ref{FigStellarD2}, but changing the disc mass (and, therefore, $\Sigma$) to $M_{d} = 5.1 \times 10^{10}$ and $M_{d} = 1.25 \times 10^{10} \ M_{\odot}$. The thickness of the resulting stellar discs is shown, as a function of $\epsilon_g$, in the bottom panel of Fig.~\ref{FigStellarDSoft}. The transition between ``thin'' and ``thick'' stellar discs clearly occurs when the softening becomes smaller or larger than the critical value of equation (\ref{EqEpsCrit}), which is indicated by the vertical arrows.

The discussion above provides some guidance to the choice of numerical parameters in numerical simulations. Resolving the vertical structure of unstable discs will inevitably lead to the  formation of large numbers of dense, tightly-bound clumps of gas that will rapidly turn into stars, leading to unacceptably thick stellar discs as the clumps later merge and disrupt. One way of preventing this outcome is by choosing feedback algorithms that strongly regulate and limit the star formation efficiency in each clump, allowing only a small fraction of its mass to be transformed into stars before the remaining gas is effectively dispersed by feedback. The newly formed stars would no longer be self-bound and should be quickly mixed within the disc by differential rotation, limiting the kinematic heating of the disc.

\section{Summary and Conclusions}
\label{SecConc}

We study the vertical structure and stability of centrifugally-supported polytropic gaseous discs embedded in cold dark matter (CDM) halos, and contrast analytical results with those of numerical techniques used in some of the latest cosmological hydrodynamical simulations. The aim of this comparison is not only to validate the numerical techniques but also  to assess the effects of limited numerical resolution, such as gravitational softening or a finite number of particles. 

We use the {\tt Gadget-2} SPH code, modified to include the gravitational acceleration of a CDM halo (modeled as a rigid, spherical  Navarro-Frenk-White potential), and adopt the same polytropic equation of state, $P\propto \rho^\Gamma$, adopted by the EAGLE suite of cosmological simulations ($\Gamma=4/3$).

The equilibrium vertical disc structure is set by the balance between the gas pressure and the compressive forces of the halo and of the disc's self-gravity. We distinguish between two regimes, when either the halo dominates (non-self-gravitating, NSG discs), or the disc dominates (self-gravitating, SG discs). 

Our main conclusions may be summarized as follows.

\begin{itemize}

\item At given radius R, the characteristic scaleheight of a disc is generally smaller than that expected either in the NSG case, where $z_{\rm H}\propto (c_s/V_c)R$, or the SG case, where $z_{\rm H}\propto c_s^2/G\Sigma$.  The mean square harmonic of the two provides a simple yet reasonably accurate estimate (equation~\ref{EqZh}).

\item The {\it shape} of the vertical density profile at given $R$ is a strong function of the polytropic index $\Gamma$, but depends only weakly on whether the disc is self-gravitating or not.

\item SG discs are generally Toomre unstable and quickly break into clumps; NSG discs are stable. The characteristic lengthscale of instabilities ($\lambda_{\rm crit}$; equation~\ref{EqLambdaCritDef}) is typically much larger than the disc scaleheight.

\item SPH simulations of gaseous exponential discs recover the expected disc scaleheight and vertical dependence quite accurately, even with as few as a few hundred particles per disc.

\item The gravitational softening of disc particles plays {\it no role} in the vertical structure of NSG discs, but may lead to artificial thickening of SG discs when the softening is comparable or larger than the expected SG scaleheight. Quantitatively, the effect of the softening depends on the difference between NSG and SG scaleheights: for softenings large enough, all discs converge to the NSG solution.

\item Large gravitational softenings may artificially stabilize otherwise unstable discs when the critical instability lengthscale is not properly resolved (i.e., for $\epsilon_g>\lambda_{\rm crit}$). Since generally $\lambda_{\rm crit}>z_{\rm H}$, discs that are well resolved vertically also adequately resolve the onset of radial instabilities.

\item Unstable discs where a large fraction of stars form in tightly self-bound clumps may be substantially thickened by the subsequent dispersal of stars that result from the merging of the clumps. Feedback mechanisms might be able to prevent this if they lead to the dissolution of the clumps before most of their mass is turned into stars.

\item Realistic galaxy discs in $\Lambda$CDM  (i.e., those inhabiting halos that follow abundance matching constraints and whose sizes are comparable to those observed, at a given galaxy mass) are expected to be NSG at low masses but SG at high masses for the EAGLE polytropic ($\Gamma=4/3$) equation of state.

\item Realistic discs in $\Lambda$CDM  have approximately flat circular velocity curves and declining density profiles. They are generally expected to ``flare'' outwards.

\item Realistic discs in $\Lambda$CDM  have surface densities that scale as $\Sigma\propto M^{1/3}$, which implies that the typical scaleheight is a very weak function of mass ($z_H\propto M^{-1/6}$) for massive, SG discs modelled with an EAGLE-like EoS. This means that a single choice of gravitational softening ($\epsilon_g\sim 100$ pc) is enough to resolve the characteristic thickness of essentially all EAGLE discs, independently of mass. Choosing much smaller values would lead to no further improvements in gas disc scaleheights.

\end{itemize}

We conclude that numerical hydrodynamical methods such as that adopted in {\tt Gadget-2} adequately reproduce the basic vertical structure of gaseous discs formed in $\Lambda$CDM cosmological simulations. If chosen carefully, numerical parameters such as the gravitational softening introduce no spurious effect on the expected thickness of polytropic discs.  Our results provide clear guidance as to how to choose these parameters for a given effective equation of state; we provide specific recommendations in Sec.~\ref{SecSoftening}. For example, choosing the softening so that the vertical structure of the least massive disc that is expected to be self-gravitating in a cosmological simulation is properly resolved should be enough to resolve the vertical structure of discs of all masses in $\Lambda$CDM adequately. For the EAGLE EoS a gravitational softening of order $\sim 100$ pc should be enough.

We close by noting that our simulation series does not consider two further effects that should be of importance for the vertical structure of simulated discs. One of them is numerical, and concerns the ``noise'' in the gravitational potential introduced when the dark halo is modeled with a ``live'' ensemble of particles. The second concerns the effects of energetic feedback on the gaseous disc, which may introduce wind-driven bubbles and bulk motions in the gas that would add to the thermal pressure and increase the disc's thickness. The two effects are best studied together in a controlled suite of simulations like the one we present here, but including realistic initial conditions and a suitably calibrated feedback module. We plan to present results of such experiments in future work.

\section*{Acknowledgements}
We acknowledge useful discussions with Pablo Ben\'itez-Llambay, Tom Theuns and Richard Bower. We also thank the EAGLE board for allowing us to use a simplified version of the {\tt EAGLE} code to run the simulations presented in Sec.~\ref{SecStellarD}.
We have benefited from the following public {\tt PYTHON} packages: {\tt NUMPY}
~\citep{Van2011numpy}, {\tt SCIPY}~\citep{Jones2001}, {\tt MATPLOTLIB}~\citep{Hunter2007}, {\tt IPYTHON}~\citep{Perez2007} and {\tt PY-SPHVIEWER}~\citep{Benitez-Llambay2015}. This work was supported by the Science and Technology Facilities Council (grant number ST/L00075X/1) and the European Research Council (grant numbers GA 267291 ``Cosmiway''). ADL acknowledges financial support from a COFUND Junior Research Fellowship, and from a Future Fellowship from the Australian Research Council (project number FT160100250). This work used the DiRAC Data Centric system at Durham University operated by the Institute for Computational Cosmology on behalf of the STFC DiRAC HPC Facility (www.dirac.ac.uk). This equipment was funded by BIS National E-infrastructure capital grant ST/K00042X/1, STFC capital grants ST/H008519/1 and ST/K00087X/1, STFC DiRAC Operations grant ST/K003267/1 and Durham University. DiRAC is part of the National E-Infrastructure.

\bibliographystyle{mnras}
\bibliography{my_biblio}

\appendix

\section{Vertical density profile of self-gravitating polytropic disc}
\label{App:Appendix_density_profiles}

Here we show that the vertical density profile derived for a NSG polytropic disc constitutes a good approximation to the density profile of a SG disc. The vertical density profile can be derived by solving the hydrostatic equilibrium equation,
\begin{equation}
\label{App:hydro_equi}
\displaystyle\frac{1}{\rho_g} \displaystyle\frac{\partial P}{\partial z} = -\displaystyle\frac{\partial \Phi_g}{\partial z},
\end{equation}
in which $\rho_g$ is the local gas density of the disc, $P$ is the pressure, related to the density though a polytropic equation of state $P = P_{\rm eos} (\rho_{g}/\rho_{\rm eos})^{\Gamma}$, with $P_{\rm eos}$ and $\rho_{\rm eos}$ being constants that determine the normalization of the relation, and $\Gamma$ is the polytropic index. The gravitational potential sourced by the disc, $\Phi_{g}$, is related to the mass distribution of the system through the Poisson equation,
\begin{equation}
\label{App:Poission}
\displaystyle\frac{\partial^2 \Phi_{g}}{\partial z^2} = 4 \pi G \rho_{g}.
\end{equation}
Differentiating equation (\ref{App:hydro_equi}) with respect to $z$, writing the pressure of the system in terms of the density, and using equation (\ref{App:Poission}), we obtain the following differential equation for the density profile,
\begin{equation}
\label{App:Density_equation}
\displaystyle\frac{\partial}{\partial \tilde z} \left ( \tilde \rho_{g}^{\Gamma -2} \displaystyle\frac{\partial \tilde \rho_{g}}{\partial \tilde z} \right ) + \tilde \rho_{g} = 0,
\end{equation}
where the normalized variables, $\tilde \rho_g$ and $\tilde z$, are defined by:
\begin{equation}
\begin{cases}
\rho_g = \tilde \rho_{g} \rho_{\rm eos} \\
z = \tilde z \left ( \displaystyle\frac{\tilde c_s^2}{G\rho_{\rm eos}} \displaystyle\frac{\Gamma}{4\pi} \right )^{1/2} \\
\tilde c_s^2 = P_{\rm eos}/\rho_{\rm eos}.
\end{cases}
\end{equation}
As noted by~\cite{Goldreich1965}, equation (\ref{App:Density_equation}) has simple solutions for particular values of $\Gamma$. For $\Gamma=1$ and $\Gamma=2$ the solutions are:
\begin{equation}
\displaystyle\frac{\rho_{g}(R,z)}{\rho_{g}(R,0)} = \begin{cases}
\sech^2 \left [ \left ( \displaystyle\frac{2\pi G \rho_{g}(R,0)}{c_{s,0}^2} \right )^{1/2} z \right ],  &\mbox{if } 	\Gamma =1 \\
\cos \left [ \left ( \displaystyle\frac{2\pi G \rho_{\rm eos}^2}{P_{\rm eos}} \right )^{1/2} z \right ], &\mbox{if } \Gamma = 2, \\
\end{cases}
\end{equation}
where $\rho_{g}(R,0)$ and $P(R,0)$ are the density and pressure at the midplane of the disc, respectively.
Similarly to equation (\ref{EqNSGRhoZ}), equation (\ref{App:Density_selfgrav}) enables us to define the scaleheight parameter of a SG polytropic disc for $\Gamma = 1$ and $\Gamma = 2$:
\begin{equation}
H_{SG}(R) = \begin{cases}
\left ( \displaystyle\frac{c_{s,0}^2}{2\pi G \rho_{g}(R,0)} \right )^{1/2}, &\mbox{if } \Gamma = 1. \\
\displaystyle\frac{\pi}{2} \left ( \displaystyle\frac{P_{\rm eos}}{2\pi G \rho_{\rm eos}^2} \right )^{1/2}, &\mbox{if } \Gamma = 2. \\
\end{cases}
\end{equation}
Previous solutions share a number of similarities when contrasted with equation (\ref{Eq:Height-nograv}). Indeed, for $\Gamma = 1$, the scaleheight parameter, $H_{SG}$, is defined as the $z$-coordinate above (or below) which the density drops by a factor $\sech^2(1) \sim 0.42$ (recall that it is $1/e \sim 0.37$ for a NSG isothermal disc). Moreover, for a SG isothermal disc the density (and pressure) vanish at infinity, similarly to what we found in Section~\ref{SecNSG} for a NSG disc. For $\Gamma = 2$, the disc is effectively polytropic. In this case, the density vanishes at a finite height, thus defining the ``true'' height of the disc, similarly to what we found for NSG polytropic discs. We now compare the value of the scaleheight parameter given by equation (\ref{App:heights}), $H_{NSG}$, to $H_{SG}$ (equation \ref{Eq:Height_grav}):

\begin{equation}
\label{App:heights_comparison}
\displaystyle\frac{H_{NSG}}{H_{SG}}= \begin{cases}
 F_c = \displaystyle\frac{\sqrt{\pi}}{2} \sim 0.9, &\mbox{if } \Gamma = 1. \\
 \displaystyle\frac{\pi}{2}F_c \sim 1.1,  &\mbox{if } \Gamma = 2, \\
\end{cases}
\end{equation}
where $F_c$ is given by equation (\ref{Eq:Correction_factor}). Thus, we conclude that the approximate scaleheight parameter derived in Sec.~\ref{SecNSG} (equation~\ref{Eq:Height_grav}) differ by $\sim 10 \%$ relative to the actual scaleheight parameter, which is acceptable for our purposes.

\section{Local stability of SG discs and the impact of the gravitational softening}
\label{App:Instabilities}
The dispersion relation for axisymmetric disturbances in an (infinitesimally thin) rotating gas disc reads\footnote{See, e.g., Sec. 6.2.3 in \cite{Binney2008}.}:
\begin{equation}
 \omega^2 = \kappa^2 - 2\pi G \Sigma k + c_s^2 k^2,
\end{equation}
where $\omega$ and $k$ are the frequency and the wavenumber of the perturbation, respectively. The gas disc is unstable if $\omega^2<0$, which happens when $Q=\kappa c_s / \pi G \Sigma < 1$. The line of neutral stability of the system is given by:
\begin{equation}
 \kappa^2 - 2\pi G \Sigma k + c_s^2 k^2 = 0,
\end{equation}
which can be written in terms of $Q$ as follows:
\begin{equation}
 1-\left (\displaystyle\frac{k}{k_{\rm crit}} \right ) + \left (\displaystyle\frac{Q}{2} \right )^2  \left (\displaystyle\frac{k}{k_{\rm crit}} \right )^2 = 0,
\end{equation}
with $k_{\rm crit} = \kappa^2 / 2\pi G \Sigma$.
Thus, the critical value, $Q_{c}$, below which a perturbation with wavenumber $k$ is unstable is
\begin{equation}
 Q_{c}(k) = 2 \sqrt { \displaystyle\frac{k_{\rm crit}}{k} - \left ( \displaystyle\frac{k_{\rm crit}}{k} \right )^2 },
\end{equation}
or, equivalently, in terms on the perturbation's wavelength $\lambda = 2\pi/k$,
\begin{equation}
\label{Eq:Toomre_wavelength}
 Q_{c}(\lambda) = 2 \sqrt { \displaystyle\frac{\lambda}{\lambda_{\rm crit}} - \left ( \displaystyle\frac{\lambda}{\lambda_{\rm crit}} \right )^2 }
\end{equation}
Perturbations with $\lambda > \lambda_{\rm crit}$ and $\lambda << \lambda_{\rm crit}$ are stable. The ``most'' unstable wavelength is $\lambda_{\rm crit}/2$, which correspond to a value $Q_{\rm c} = 1$. 

In order to resolve instabilities of wavelength $\lambda \sim \lambda_{\rm crit}/2 = 2\pi^2G\Sigma/\kappa^2$, gravity must be properly resolved on scales smaller than $\lambda_{\rm crit}$; otherwise, artificial stabilization may occur. This means that gravity must not be softened on scales $h_{\epsilon} \sim 3\epsilon_{g} \lesssim \lambda_{\rm crit}$, where $\epsilon_{g}$ is the ``equivalent'' Plummer softening, which implies:
\begin{equation}
 \epsilon \lesssim \epsilon_{\rm crit} = \displaystyle\frac{2}{3}\displaystyle\frac{\pi^2G\Sigma}{\kappa^2} \sim (0.58 \rm \ kpc) \left (\displaystyle\frac{\Sigma}{10^8 M_\odot \rm kpc^{-3}} \right ) \left ( \displaystyle\frac{\kappa^2}{70 \rm \ km \ s^{-1} kpc^{-1}} \right )^2. 
\end{equation}
We have verified that this equation is a good approximation to the critical softening below which radial instabilities are resolved. See Sec.~\ref{SecSoftInst}.

\section{Constructing a general solution to the scaleheight parameter}
\label{App:general_scaleheight}
In Sec.~\ref{SecPrelim} we stated that a disc is non-self-gravitating if the vertical acceleration profile is primarily described by equation (\ref{Eq:Non-self-gravitating-intro}), and self-gravitating if it is largely described by equation (\ref{Eq:Self-gravitating-intro}). Here we calculate more precisely the transition between these two regimes and justify why the square harmonic mean (equation~\ref{EqZh}) constitutes a good approximation to the scaleheight parameter of polytropic discs. 

The condition that determines whether a disc is self-gravitating or non-self-gravitating can be obtained by comparing the contribution of the disc and the halo to the vertical gravitational acceleration at $(R,z)=(R,H)$:
\begin{equation}
\label{Eq:Self-gravity_criterion}
 \displaystyle\frac{\left ( \partial \Phi_{h}/\partial z \right ) }{ \left ( \partial \Phi_{g}/\partial z \right ) }(R,H) = \displaystyle\frac{V_c^2(R)/R}{2\pi G \Sigma(R)} \left ( \displaystyle\frac{H}{R} \right ) \equiv \displaystyle\frac{1}{3 F_c} \left [ \displaystyle\frac{\bar \rho_{\rm dm} (R)}{\rho_{g}(R,0)(R)} \right ],
\end{equation}
in which $\bar \rho_{\rm dm} (R) = M_{\rm dm}(<R)/(4/3\pi R^3)$ is the mean enclosed dark matter density at a given radius, related to the circular velocity of the system by $(V_c/R)^2 = (4/3) \pi G \bar \rho_{\rm dm}(R)$, in a dark matter-dominated system; $\rho_{g}(R,0)$ is the midplane density of the disc, and $F_c$ is given by equation (\ref{Eq:Correction_factor}). In terms of the actual densities, the system is either SG or NSG at $R$ according to the following criterion:
\begin{equation}
\label{Eq:Self-gravitating_appendix}
\begin{cases}
\bar \rho_{\rm dm} (R) \, \ll 3 F_{c} \rho_{g}(R,0) &\mbox{if the disc is SG}, \\
\bar \rho_{\rm dm} (R) \, \gg 3 F_{c} \rho_{g}(R,0) &\mbox{if the disc is NSG},\\
\end{cases}
\end{equation}
which suggests that at a given radius, the parameter that determines whether a disc is SG or not is the ratio between the mean enclosed dark matter density, $\bar \rho_{\rm dm}(R)$, and the local gas density at the midplane, $\rho_{g}(R,0)$. 

We now propose a formula that converges to the desired scaleheight parameter in the asymptotic SG and NSG regimes, with the property of transitioning between these two regimes according to criterion (\ref{Eq:Self-gravitating_appendix}):
\begin{equation}
\label{Eq:General_height}
H(R) = H_{NSG}(R) \left ( 1 + \displaystyle\frac{\rho_g(R,0)}{\beta(\Gamma) \bar \rho_{\rm dm} } \right) ^{-1/2}                 
\end{equation}
where $H_{NSG}(R)$ is given by equation (\ref{Eq:Height-nograv}) and $\beta(\Gamma)$, defined by,
\begin{equation}
\label{Eq:beta_general}
\beta(\Gamma) = \displaystyle\frac{2}{3 \Gamma  F_c^2 \alpha^2(\Gamma)},
\end{equation}
ensures that the scaleheight parameter is $H_{SG}$ when the disc is self-gravitating. Recall that the midplane density of the disc is related to the surface density through equation (\ref{Eq:rho_to_sigma}).

It is straightforward to see that equation (\ref{Eq:General_height}) converges to the desired solution in the asymptotic cases. Indeed, in the limit $\rho_g(R,0) << \beta \bar \rho_{\rm dm} \le 3 F_c$, the disc is effectively non-self-gravitating, and the solution for $H_{NSG}$ (equation \ref{Eq:Height-nograv}) is recovered.  On the other hand, if the disc is SG, $\rho_g(R,0) \gg 3 F_c \bar \rho_{\rm dm} \ge \beta \bar \rho_{\rm dm}$, and the solution for $H_{SG}$ (equation~\ref{Eq:Height_grav}) is recovered. We have also validated equation (\ref{Eq:General_height}) at intermediate regimes, where either the gravity of the halo and the disc are both important in determining the disc's scaleheight, and found that it constitutes a very accurate description of the detailed scaleheight of polytropic discs. Note, however, that equation (\ref{Eq:General_height}) is inconvenient, as the solution is directly expressed in terms of the midplane density of the disc, which, for a given surface density, depends explicitly on the scaleheight parameter (see equation~\ref{Eq:rho_to_sigma}). Thus, equation (\ref{Eq:General_height}) defines the scaleheight parameter of a polytropic disc implicitly. It is possible to derive an equivalent (and perhaps more convenient) formula by studying the asymptotic behaviour of equation (\ref{Eq:General_height}). Indeed, it is straightforward to see that equation (\ref{Eq:General_height}) approaches the SG regime as:
\begin{equation}
\label{Eq:convergence_sg}
H^2 = H_{NSG}^2 \left [ 1+ \left ( \displaystyle\frac{\eta}{Q} \right )^2 \right ]^{-1} = H_{NSG}^2 \left [ 1+ \left ( \displaystyle\frac{H_{NSG}}{H_{SG}} \right )^2 \right ]^{-1},
\end{equation}
where we have used equation (\ref{Eq:rho_to_sigma}) to relate the midplane density of the disc to the surface density, and also assumed that the epicyclic frequency of the disc is $\kappa = 2 V_c/R$; $Q$ is the Toomre parameter and $\eta$ is:
\begin{equation}
\label{Eq:Eta}
\eta = 2 \alpha(\Gamma) \Gamma F_c.
\end{equation}
 Similarly, equation (\ref{Eq:General_height}) approaches the NSG regime as:
\begin{equation}
\label{Eq:convergence_nsg}
H^2 = H_{NSG}^2 \left [ 1+ \displaystyle\frac{\eta}{Q}  \right ]^{-1} = H_{NSG}^2 \left [ 1+ \left ( \displaystyle\frac{H_{NSG}}{H_{SG}} \right ) \right ]^{-1},
\end{equation}
A convenient function that captures these properties is, in fact, the square harmonic mean:
\begin{eqnarray}
\label{Eq:hamonic_mean_1}
H^2 &=& \displaystyle\frac{H_{NSG}^2}{1+\frac{H_{NSG}}{H_{SG}} \left ( 1+\frac{H_{NSG}}{H_{SG}} \right ) } \\ \nonumber 
&=& \left ( \displaystyle\frac{1}{H_{NSG}^2}+\displaystyle\frac{1}{H_{NSG}H_{SG}}+\displaystyle\frac{1}{H_{SG}^2} \right )^{-1}
\end{eqnarray}

Inspired by this, we propose the following equation, which proved to be as accurate as equation (\ref{Eq:General_height}) in recovering the scaleheight parameter of polytropic discs:
\begin{equation}
\label{Eq:harmonic_mean_2}
H^2 = \left ( \displaystyle\frac{1}{H_{NSG}^2}+\displaystyle\frac{1}{2 H_{NSG}H_{SG}}+\displaystyle\frac{1}{H_{SG}^2} \right )^{-1}
\end{equation}
Equation (\ref{Eq:General_height}) is particularly useful for taking into account the impact of the gravitational softening on the scaleheight parameter. Indeed, since it takes into account the transition between the NSG and the SG regimes naturally, the impact of the gravitational softening may be incorporated by adding a correction function to the SG term:
\begin{equation}
\label{Eq:General_height_softened}
H(R) = H_{NSG} \left [ 1 + \displaystyle\frac{\xi(\epsilon_{g})}{\beta(\Gamma)} \displaystyle\frac{\rho_{g}(R,0)}{\bar \rho_{\rm dm}} \right] ^{-1/2},
\end{equation}
where $0 \le \xi(\epsilon) \le 1$ is the correction function of the form:
\begin{equation}
 \xi(\epsilon_{g}) = \displaystyle\frac{1}{1+\left (\epsilon_{g}/H_{SG} \right)^{\nu}},
\label{Eq:Correction_function}
\end{equation}
in which $\nu$ must be fitted from numerical experiments. For the example shown in Fig.~\ref{FigZEps}, we find $\nu \sim 1.4$ yields good results. Equation (\ref{Eq:General_height_softened}) enables us to derive the ``softening-corrected'' version of equation (\ref{Eq:harmonic_mean_2}):
\begin{equation}
\label{Eq:harmonic_mean_2_soft}
H^2 = \left ( \displaystyle\frac{1}{H_{NSG}^2}+\displaystyle\frac{\xi(\epsilon_g)}{2 H_{NSG}H_{SG}}+\displaystyle\frac{\xi(\epsilon_g)}{H_{SG}^2} \right )^{-1}
\end{equation}
Thin dashed lines in Fig.~\ref{FigZEps} show the characteristic scaleheight of ``softened'' polytropic discs, contrasted with numerical experiments. Clearly, this simple correction provides an excellent quantitative description of the artificial thickening caused by the use of a finite gravitational softening.

\section{Discs become self-gravitating before going unstable}
\label{Subsec:self-gravity}

The onset of instabilities in the plane of the disc is well predicted by the dimensionless Toomre parameter, $Q$,~\citep{Toomre1964}, defined as:
\begin{equation}
Q = \displaystyle\frac{c_s \kappa}{\pi G \Sigma},
\end{equation}
\noindent where $\kappa $ is the epicyclic frequency, related to the angular velocity, $\Omega$, by:
\begin{equation}
\kappa^2 = \displaystyle\frac{2\Omega}{R} \displaystyle\frac{d}{dR} \left (R^2 \Omega \right ).
\end{equation}
A disc will become Toomre unstable if $Q$ drops below $Q_{\rm crit} \approx 1$. A Toomre unstable disc is in general self-gravitating, but not the other way around. Indeed, when inserting equation (\ref{Eq:Height_grav}) into equation (\ref{Eq:Self-gravitating_appendix}) we see that the condition for a disc to be self-gravitating is:
\begin{equation}
\label{Eq:Self-gravity-Toomre}
  \displaystyle\frac{(V_c/R) c_{s,0}}{\pi G \Sigma} = Q \lesssim \left ( 8 \Gamma F_c \right )^{1/2},
\end{equation}
in which we assumed that $\Omega$ is a weak function of $R$, so that $\kappa = 2\Omega = 2V_c/R$. The factor $\left ( 8 \Gamma F_c \right )^{1/2}$ ranges between $2$ and $3$ for usual values of $\Gamma$. Our numerical experiments suggest that galaxies become unstable for values of $Q \lesssim 0.6 \lesssim \left ( 8 \Gamma F_c \right )^{1/2}$~\citep[see][for a similar result]{Wang2010}. Thus, we conclude that self-gravitating polytropic discs are not necessarily Toomre unstable, at least within a narrow range of surface densities.

\end{document}